\documentclass[aps, superscriptaddress,  twocolumn]{revtex4}
\usepackage{graphicx}% Include figure files

\usepackage{dcolumn}% Align table columns on decimal point
\usepackage[mathlines]{lineno}
\usepackage{physics}
\usepackage{hyperref}
\usepackage{multirow}
\usepackage{bm}% bold math
\usepackage{epstopdf}
\usepackage{epsfig}
\usepackage{ulem}
\usepackage{bbold}
\usepackage[dvipsnames]{xcolor}

\usepackage{amsmath}
\begin{document}
%%%%%%%%%%%%%%%%%%%title%%%%%%%%%%%%%%%%%%%
\title{Topological space-time waves in complex channels}

\author{Renwei Zou}
\thanks{These authors contributed equally to this work.}
\affiliation{College of Optical Science and Engineering, Zhejiang University, Hangzhou 310058, China}

\author{Kelsey Everts}
\thanks{These authors contributed equally to this work.}
\affiliation{College of Optical Science and Engineering, Zhejiang University, Hangzhou 310058, China}
\affiliation{School of Physics, University of the Witwatersrand, Private Bag 3, Wits 2050, South Africa}

\author{Andrew Forbes}
%\email[email:]{andrew.forbes@wits.ac.za}
\affiliation{College of Optical Science and Engineering, Zhejiang University, Hangzhou 310058, China}
\affiliation{School of Physics, University of the Witwatersrand, Private Bag 3, Wits 2050, South Africa}

\author{Yungui Ma}
\email[email:]{yungui@zju.edu.cn}
\affiliation{College of Optical Science and Engineering, Zhejiang University, Hangzhou 310058, China}
\affiliation{Jiangxi Qiushi Advanced Research Institute, Nanchang City 330038, Jiangxi Province, China}
%\email[Corresponding author: ]{yungui@zju.edu.cn}

\date{\today}

%%%%%%%%%%%%%%%%%%%%%abstract%%%%%%%%%%%%%%%%%%%%
\begin{abstract}
\noindent \textbf{Structured light in space and time has become a powerful playground in which to explore the fundamental physics of wave systems, while simultaneously introducing new exotic forms of light, from spatio-temporal vortices to toroidal pulses of light.  Yet their creation remains restricted by complex optical systems while directly observing their evolution in arbitrary channels remains elusive, complicated by the interplay of dispersion and diffraction and exacerbated by the lack of suitable detection tools. Here we create topological space-time beams in a single step by a resonant response of a symmetry-broken metasurface, mixing spatial, temporal and polarisation degrees of freedom in a single microwave field. Our realisation in the microwave regime allows us to directly observe their dynamics in arbitrary channels, from free-space to complex random media, showing the preservation of topology while the foundational degrees of freedom are scrambled.  We use our control to show how to unscramble the space-time properties for the first crosstalk-free transmission of space-time waves. Our work advances the physics of space-time waves, introduces a new toolkit for their creation and control, and offers an exciting roadmap to their exploitation in real-world scenarios, e.g., for robust communications through noisy channels.}
\end{abstract}

\maketitle

\section*{Introduction}

\noindent Structuring light's degrees of freedom (DoFs) \cite{he2022towards,forbes2021structured,forbes2025progress}, has become highly topical of late, revolutionizing the way we approach the tailoring of light for probes of fundamental physics and applications alike. Leveraging both space and time simultaneously has seen explosion of interest recently \cite{forbes2025structured,zhan2024spatiotemporal,liu2024spatiotemporal,yessenov2022space}, with control extending to spatio-temporal vortices (STOVs) \cite{jhajj2016spatiotemporal,hancock2019free,chong2020generation,bliokh2021spatiotemporal}, toroidal pulses of light \cite{wan2022toroidal,zdagkas2022observation}, spatiotemporal vectorial beams (STVBs) \cite{huang2025transverse,Teng2025construction,pires2025structuring,zhang2026extending} and generally the control of transverse orbital angular momentum (OAM) \cite{wang2021engineering,zang2022spatiotemporal}. These exotic forms of light are driving exciting applications, from analog computation \cite{huang2022spatiotemporal,zhou2023electromagnetic,huang2026experimental,huang2025arithmetic} to information transmission and communications \cite{cao2023propagation,huang2024spatiotemporal,feng2025spatiotemporal,kupferman2026perfect}, made possible by a sophisticated toolkit ranging from bulk pulse shapers \cite{cao2022vectorial} to compact nanophotonic implementations \cite{ni2024three,liu2024exploiting}. 

These demonstrations have remained largely confined to optical frequencies, where direct observation of spatiotemporal dynamics is still challenging. An exciting prospect is the use of metasurface-enabled STVB generation to push the wavelength range to regions that allow direct observation of the space-time properties, e.g., the microwave regime. Unfortunately current metasurface functionality has been limited to scalar spatiotemporal vortex beams with a single polarization state \cite{huo2024observation,zhou2025quasi}, restrained by the distinct technical hurdle of integrating dispersive spectral modulation with spatial wavefront shaping for each orthogonal polarization component within a single device, prohibiting the creation of vectorial and topological space-time waves.

\begin{figure*}[hpt!]
	\includegraphics[width=0.9\textwidth]{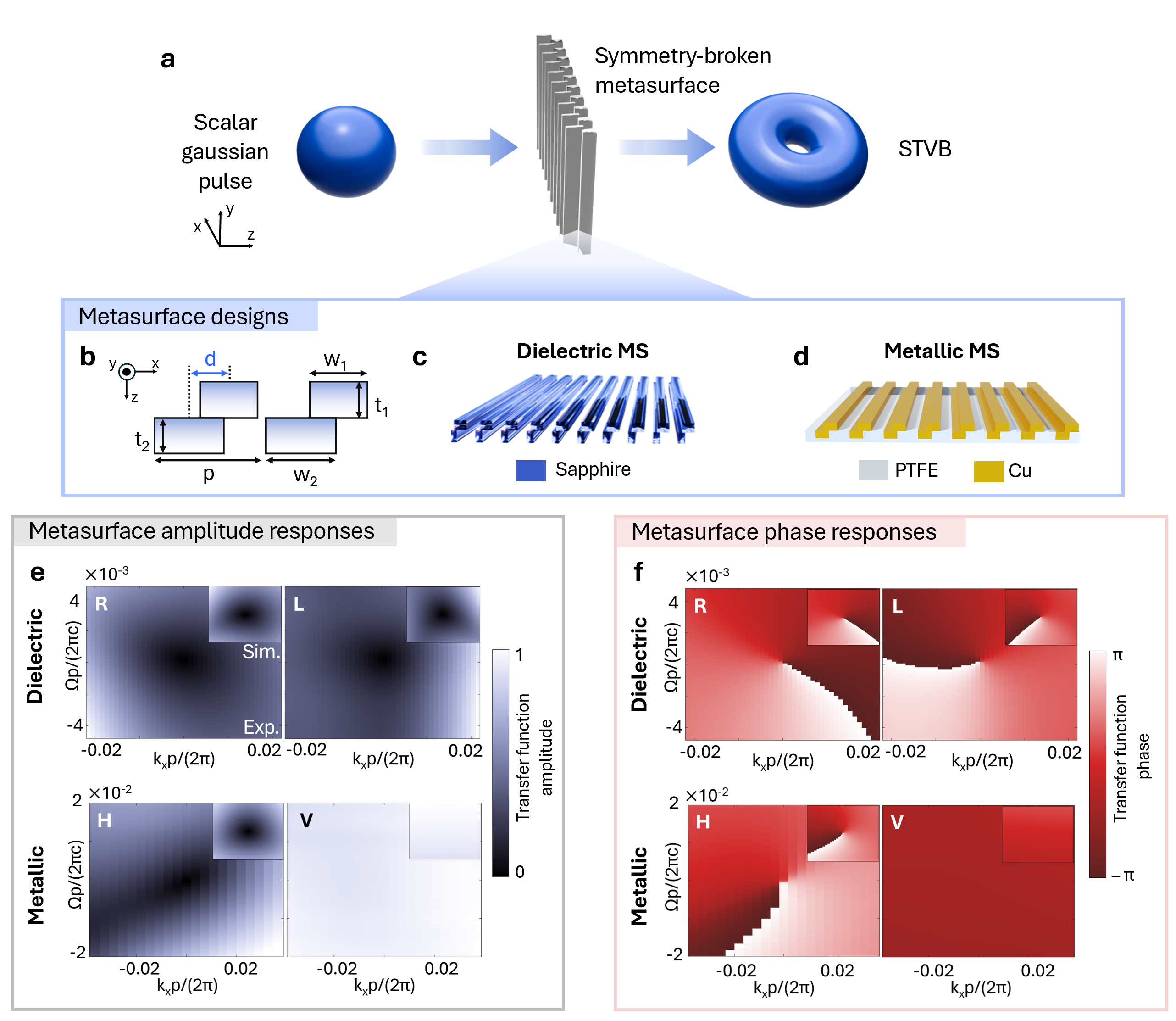}%{Figures_new/Fig1_v5.png}
	\caption{\textbf{Metasurface assisted STVB generation.} \textbf{a} Illustration showing STVBs generated with symmetry broken subwavelength grating MS. When a scalar Gaussian pulse is incident on our sub-wavelength grating metasurface, the beam's structure is modulated to form a vortex on one or more polarisations generating an output STVB.  Intensity isosurfaces show input and output beams for STVB with $|\ell|=1$ on both polarisation components. \textbf{b} Two dimensional cross-section of the symmetry broken grating MS consisting of two offset gratings with bar thicknesses $t_{1,2}$ and widths $w_{1,2}$ with a common period $p$ with the upper and lower gratings laterally displaced by dislocation distance $d$. \textbf{c} Three-dimensional rendering of the dielectric sapphire MS and \textbf{d} the metallic MS consisting of poly-tetrafluoroethylene (PTFE) and copper (Cu). \textbf{e} MS amplitude transfer functions plotted in false color as a function of spatial frequency $k_x$, period $p$ and frequency detuning $\Omega=\omega-\omega_0$ from centre frequency $\omega_0$. Right (R) and left (L) elliptical polarisation components shown for the dielectric MS (upper row). Horizontal (H) and vertical (V) shown for the metallic MS (lower row).  Main plots show experimental measurements and insets show simulations. \textbf{f} Corresponding phase transfer functions for the same cases shown in \textbf{e}.}
	\label{fig:concept_setup}
\end{figure*}

Here, we theoretically propose and experimentally realize novel, compact metasurface platforms that synthesizes 4D topological space-time waves in a single step. By leveraging symmetry-broken grating designs we are able to efficiently couple space, time, and polarization to generate STVBs with high fidelity and pronounced topological features, while our realisation in the microwave allows us to directly observe its space-time properties.  We leverage on this to present the first studies of space-time light through complex channels, from free-space with the interplay of dispersion and diffraction, to random media. We develop a customized measurement toolkit to investigate the complex dynamics of these waves and use it to show that the fundamental DoFs such as the amplitude, phase and polarization structure of these STVB are inevitably scrambled, but reveal how the underlying topology and classical entanglement remains intact.  We show how a new basis can be derived that unscrambles the space-time information, for crosstalk-free transmission, with direct relevance to modern day communication systems (6G and beyond) that are microwave based. Our work provides a new paradigm for the creation and measurement of STVBs and can easily be extended to optical frequencies, opening a new route for harnessing multi-dimensional structured light in space and time.

\section*{Results}
\noindent \textbf{Metasurface enabled spatiotemporal vectorial light.}  We wish to create STVBs of the form
\begin{equation}
       \ket{\psi} = \ket{0}_y \left( \ket{\ell_1}_{xt} \ket{e_1} +\ket{\ell_2}_{xt} \ket{e_2} \right) 
        \label{eq:STVB}
\end{equation}
where the polarization DoF is denoted by two orthogonal polarisations of the form $|e_1\rangle$ and $|e_2\rangle$, and $\ell_i$ is the azimuthal index of a spatio-temporal Laguerre-Gaussian state denoted by $ \ket{\ell_i}_{xt}$, with transverse-OAM (t-OAM) circulating in the $x-t$ plane. The output beam is Gaussian in $y$ for both polarizations, denoted by $\ket{0}_y$. Such an optical field has inherent coupling between its  transverse spatial $ (x,y) $ and temporal ($ t $) coordinates \cite{bekshaev2024spatiotemporal}, for a 4D STVB that is separable in the $y$ amplitude profile but non-separable in the $x-t$ and polarisation basis. When $\ell_1 = -\ell_2$ we form non-topological STVBs, while when $|\ell_1| \neq |\ell_2|$, we form topological STVBs where the non-zero topological charge difference means that the topological STVB covers all polarization states.

We employ a single metasurface (MS) solution for its construction as shown in Figure~\ref{fig:concept_setup}\textbf{a} which takes as input a tilted polarisation input Gaussian pulse which is shown as a spherical intensity isosurface on the left. The desired MS output is a STVB with t-OAM circulating perpendicular to the propagation direction (blue arrows) with a donut 3D intensity structure in each of the component STOVs as indicated by the intensity isosurface for the non-topological STVB. We demonstrate two such MS designed for operation in the microwave regime, each implementing a distinct transformation of the input pulse. We begin by describing the common underlying MS design principle for both which is rooted in the asymmetric sub-wavelength grating cross-section shown in Figure ~\ref{fig:concept_setup}\textbf{b}. This consists of two offset 1D gratings each defined by tunable structural parameters: thicknesses $t_{1,2}$ and widths $w_{1,2}$ with a common sub-wavelength period $p$ where the upper and lower gratings are laterally displaced by dislocation distance $d$ which introduces the asymmetry. The first MS is fabricated from sapphire and the second from poly-tetrafluoroethylene (PTFE) and copper (Cu) as shown by the 3D rendered images in Figure~\ref{fig:concept_setup}\textbf{c} and \textbf{d} respectively. Henceforth we denote these as \textit{dielectric MS} and \textit{metallic MS} respectively. Both MS feature the same dislocated bilayer grating geometry from Figure~\ref{fig:concept_setup}\textbf{b} with the grating varying periodically along $x$ and remaining uniform along $y$, however the differences in the symmetry breaking extent, material and geometric parameters give rise to distinct transformations of the input pulse. For the dielectric MS, we engineer a dual-channel spatiotemporal differentiator which performs temporal differentiation on one polarization component and simultaneously spatial differentiation on the orthogonal polarization. This solution is based on addressing the two polarizations simultaneously by harnessing the inherent birefringent properties of the sapphire material. The resulting polarization-selective resonances produces symmetric but opposite t-OAM on orthogonal polarisations consistent with the generation of a non-topological STVB. In the metallic MS, the birefringence is abandoned forming a single-channel spatiotemporal differentiator with modulation only performed on one polarisation. The symmetry breaking is also reduced towards a quasi-bound state in the continuum which produces asymmetric t-OAM  on orthogonal polarizations ($|\ell_1| \ne |\ell_2|$) consistent with the generation of a topological STVB. Full details of the MS design are given in the Supplementary Information.

These desired properties are confirmed by the simulated and measured transfer functions shown in Figure ~\ref{fig:concept_setup}\textbf{e} and \textbf{f} which capture the full momentum-frequency response of each MS which fully characterizes its spatio-temporal action on the input pulse. The transfer function was characterised by  known plane-wave input and by measuring the transmitted field directly with an antenna. The amplitude for the optimized fabricated dielectric MS is shown in the upper row of Figure ~\ref{fig:concept_setup}\textbf{e}. This exhibits a dark intensity null in the momentum-frequency spectrum for both left (L) and right (R) elliptical polarisations yielding the amplitude profiles required for $|\ell|=1$. This can be seen by the isolated transmission zero where all wavevector components equals zero ($k_x=0$, the $\Gamma$ point) around our microwave pulse central frequency $\omega_0 = 25$ GHz where the frequency detuning, $\Omega=\omega-\omega_0$, approaches zero. The simulated transfer functions are shown in the upper insets and closely match the measured output spectra confirming that the fabricated MS faithfully reproduces the target response. The corresponding response for the fabricated metallic MS was measured with the amplitude result shown in the lower row of Figure ~\ref{fig:concept_setup}\textbf{e}. Now, instead, the intensity null is only present in the horizontal (H) polarisation producing a similar $|\ell|=1$ signature. The spatially uniform transmission of unity for all frequencies and momenta in the vertical (V) amplitude component shows that the input V Gaussian component remains unmodulated indicative of the single-channel metallic MS action. The corresponding phase transfer functions are shown in Figure ~\ref{fig:concept_setup}\textbf{f}. The dielectric MS (upper row) exhibits two opposite handed $2\pi$ phase windings on R and L  centered at $k_x=\Omega=0$ associated with opposite handed t-OAM whereas the metallic MS (lower row) shows a $2\pi$ phase winding only on H. The V phase component remains unmodulated over the same bandwidth with an approximately flat phase and therefore carries no t-OAM. Taken together and applied to an input tilted diagonally polarised pulse, the MS amplitude and phase responses lead to a transformation rule for the dielectric MS of 
\begin{equation} \label{eq:nontopological}
    \ket{0}_{xyt}\ket{D} \overset{\text{dielectric MS}}\longrightarrow \ket{0}_y \left( \ket{1}_{xt}\ket{L} + \ket{-1}_{xt}\ket{R} \right) \, ,
\end{equation} 
for non-topological STVBs where $\ket{L}$ and $\ket{R}$ represent left and right elliptical polarisation states. The transformation rule for the metallic MS creates a topological STVBs through the transformation
\begin{equation} \label{eq:topological}
    \ket{0}_{xyt}\ket{D} \overset{\text{metallic MS}}\longrightarrow \ket{0}_y \left( \ket{1}_{xt}\ket{H} + \ket{0}_{xt}\ket{V} \right) \, 
\end{equation} 
 where $\ket{H}$ and $\ket{V}$ represent horizontal and vertical polarisation states.

\begin{figure*}[hpt!]
	\includegraphics[width=0.85\textwidth]{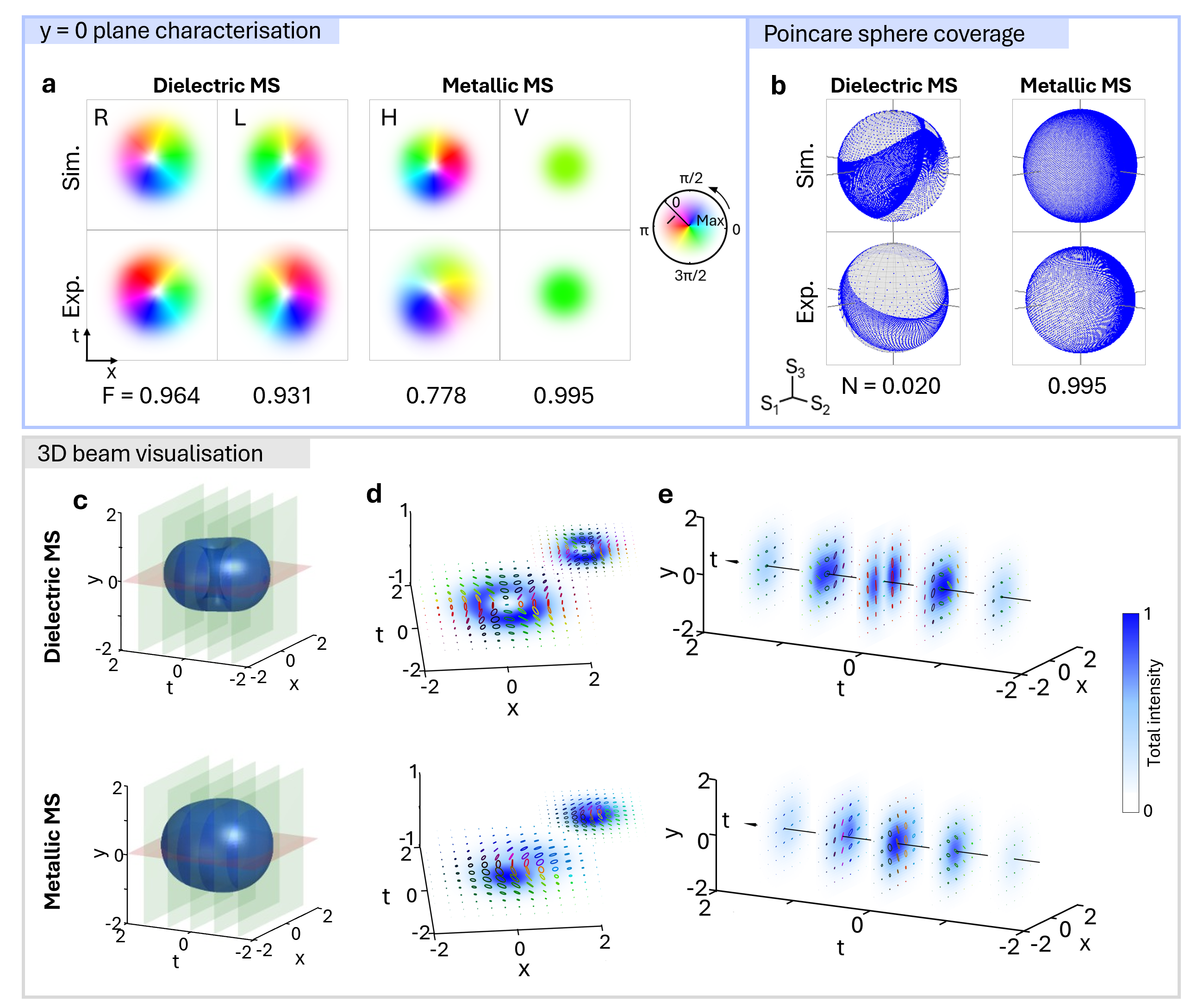}
	\caption{\textbf{Experimental realisation of meta-surface generated STVBs.} \textbf{a} Intensity and phase profiles at $y=0$ showing $x-t$ plane profiles reconstructed from the simulated and experimentally measured MS transfer functions. Component polarisation states right (R) and left (L) elliptical polarisation for the dielectric MS and horizontal (H) and vertical (V) for the metallic MS for the beam are shown. Phase is encoded by hue and intensity by brightness with fidelities (F) indicated. \textbf{b} Polarisation states of $y=0$ beam profile plotted as blue scatter points on the Poincar\'{e} sphere with experimentally measured Skyrmion wrapping number $N$ indicated for both MS. $S_i$ indicates locally renormalised Stokes parameters. \textbf{c} 3D isointensity surfaces calculated using the ideal input pulse and the measured 3D transfer functions for the dielectric MS (upper row) and metallic MS (lower row) which have been normalised by the pulse's spatial and temporal beam waists. Red and green planes correspond to 2D sliced images in \textbf{d} and \textbf{e} respectively. \textbf{d} Spatial slices at $y=0$ showing total intensity distribution with overlaid polarisation ellipses. \textbf{e} Temporal slices showing $x-y$ profiles at different times shown with overlaid polarisation ellipses. 
    }
	\label{fig:ideal_beams}
\end{figure*}

We validated these measured MS responses by experimentally reconstructing the non-topological and topological STVBs. Experimental measurement details provided in Supplementary Information. The orthogonal component polarization beam profiles for a slice of the STVB pulse at $y=0$ ($x-t$ plane) are shown in Figure~\ref{fig:ideal_beams}\textbf{a}. Intensity is given by brightness and the phase is encoded by hue. Clearly the dielectric MS (left) outputs t-OAM vortices with the characteristic donut ring-like intensity profiles in the $x-t$ plane in both R and L polarisations with opposite handed phase twists generating non-topological STVBs. On the other hand, the metallic MS only generates a t-OAM vortex on H while the V intensity profile remains Gaussian and spatially uniform flat phase. The close agreement between simulated and experimentally measured MS transfer functions and the high projection fidelities, F, listed below in Figure~\ref{fig:ideal_beams}\textbf{a} confirm the quality of these pristine STVBs and show that measurements after both MS yield the target spatiotemporal optical vortices in the $x-t$ plane. The corresponding Poincaré-sphere representations in Figure~\ref{fig:ideal_beams}\textbf{b} further distinguish the two STVBs. The blue scatter points depict the measured polarisation states from the beam's $y=0$ cross-section. $S_i$ indicates the locally renormalised Stokes parameters indicated on the Poincaré-sphere axes. The STVB generated by the dielectric MS mainly forms a broad belt/ring-like trajectory on the Poincaré sphere, consistent with a non-topological STVB which does not fully cover the sphere with the few sparse points extending further from this belt/ring originating merely from small imperfections in the MS output. The difference in ring orientation between simulation and experiment is merely due to a relative phase difference which does not change the type of STVB generated or the fidelity of the component modes. This ring corresponds to a trivial topology with a theoretical value of $N= 0$ and with a measured Poincaré sphere wrapping number of $N=0.020$. In contrast, the STVB generated by the metallic MS fully wraps around the Poincaré sphere with more scatter points covering the sphere in Figure~\ref{fig:ideal_beams}\textbf{b} uniformly once over, identifying the generated field as a spatiotemporal skyrmion with a measured skyrmion number of $N=0.995$ closely matching the theoretical value of $N=1$ at the $y=0$ plane. The full topological STVB pulse forms the expected skyrmion tube \cite{Teng2025construction} with the same behaviour along $y$ since these MS and their corresponding STVBs are $y$-independent as per Equation \ref{eq:STVB}.

We can visualize the complex structure of these 4D wave packets in several complementary ways. Figure~\ref{fig:ideal_beams}\textbf{c} shows the STVB wave packet as a three-dimensional iso-intensity surface obtained from the ideal input pulse and the measured three-dimensional transfer functions. The dielectric MS (upper row) exhibits a donut volume with a vortex singularity along $y$ at the centre of the pulse where $x=t=0$. This is because both polarisation components overlap and share a donut-shaped intensity distribution as expected from Equation \ref{eq:nontopological}. The metallic MS isosurface shown in the lower row of Figure~\ref{fig:ideal_beams}\textbf{a} lacks this central singularity because the unmodulated Gaussian component retains non-zero intensity at the pulse centre  as expected from Equation \ref{eq:topological}. Both STVBs exhibit the same structure across various $y$-planes because the pulse has an unmodulated separable Gaussian $y$-component due to the 1D nature of both MS gratings. Thus far we have focused on the intensity and phase distributions, however to show the spatiotemporal-polarization coupling more explicitly, red and green semi-transparent cross-sectional planes are indicated in Figure~\ref{fig:ideal_beams}\textbf{c} marking spatial and temporal slices. The beam cross-section at the red slice at $y=0$ is shown in Figure~\ref{fig:ideal_beams}\textbf{d}. This exhibits an inhomogeneous polarisation distribution across the $x-t$ plane depicted by the polarisation ellipses overlaid onto the beam's total intensity profile, $R+L$.  The metallic MS (lower row of Figure~\ref{fig:ideal_beams}\textbf{d}) shows that both the polarisation structure and the total intensity differ with no intensity null at the centre of the pulse due to the Gaussian component. Both STVBs are closely matched by their corresponding simulated result in the inset. Figure~\ref{fig:ideal_beams}\textbf{e} corresponds to the green slices in Figure~\ref{fig:ideal_beams}\textbf{c} and shows how the $x-y$ profile of the STVB pulse evolves with time. The black arrow indicates the pulse propagation direction (increasing time axis). Here the local polarization state is seen to evolve dynamically across the spatial envelope over time for both the topological and non-topological beams, a hallmark of 4D structured light. These transverse sliced planes thus confirm the complex spatiotemporal-polarisation structures, indicative of the non-separable nature of these waves.

\vspace{0.2cm}

\begin{figure*}[t!]
	\includegraphics[width=\linewidth]{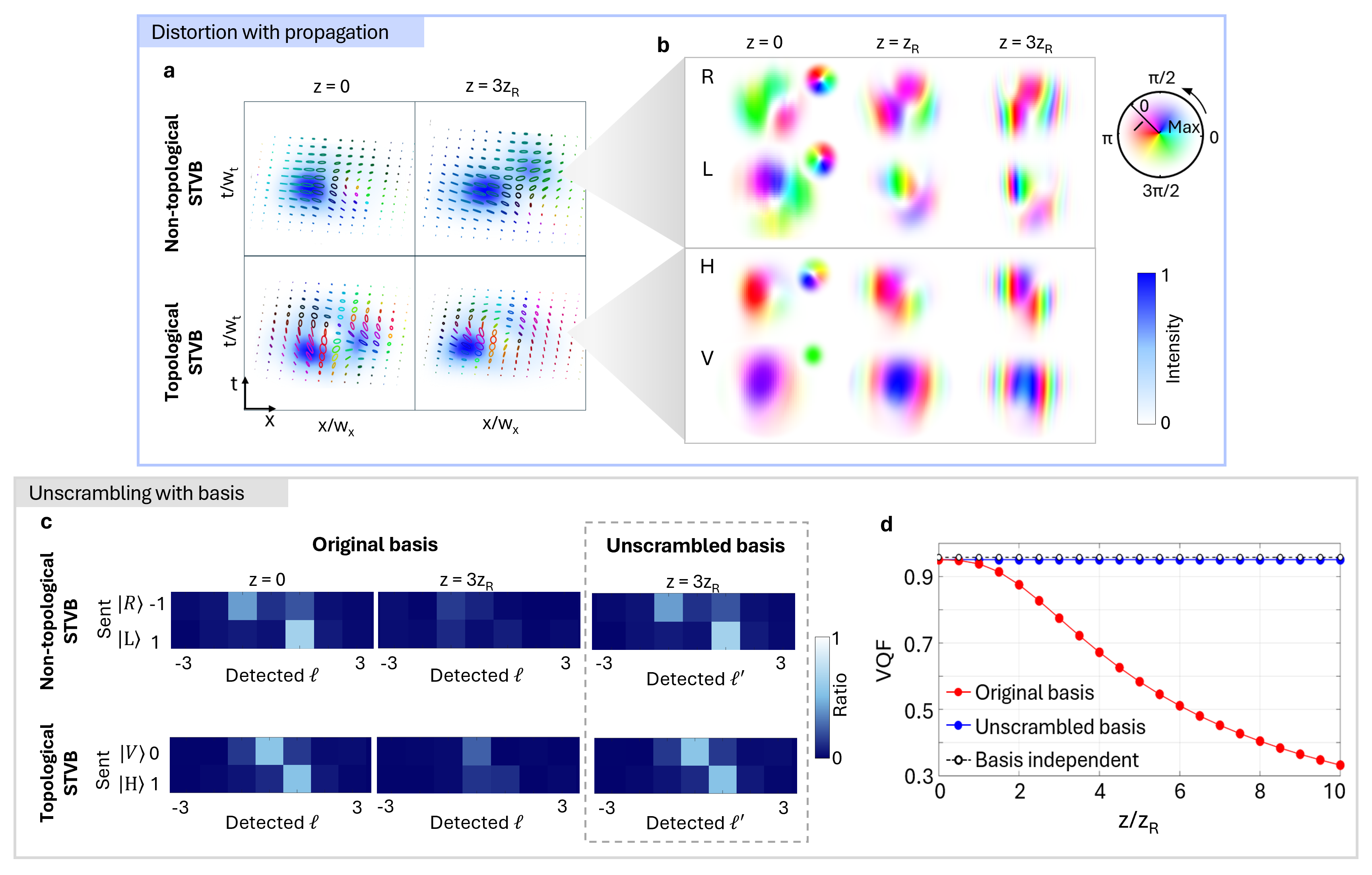}
    \caption{\textbf{Experimental results of STVBs and their free-space evolution.} 
   \textbf{a} Measured total intensity and polarization distributions of $y=0$ plane of the topological STVB and nontopological STVB at $z=0$, together with their inferred free-space evolution at $z=3z_R$ where $z_R$ is the Rayleigh range. The spatial $x$ coordinate and the temporal coordinate  have normalized by the propagated beam width $w_x(z)$, and pulse duration $w_t(z)$ respectively. \textbf{b} Measured intensity and phase distributions of the component modes of each STVB at different propagation distances. \textbf{c} Modal decomposition of each component at the $y=0$ plane at different distances where the fields are projected onto the initial spatio-temporal Laguerre-Gaussian state basis functions with different t-OAM values, $\ell$, (original basis). Projections onto the $z=3z_R$-plane adjusted detection basis $\ell'$ given by the ``unscrambled basis''. Color represents the normalized projection ratio. \textbf{d} Vectorial degree, VQF, varying with propagation distance when using the original basis, the spatiotemporal basis varying with $z$ (unscrambled basis) and the basis-independent method.}
    \label{fig:free_space}
\end{figure*}

\noindent \textbf{Evolution of STVBs in free-space.} Having established that we have the ability to generate STVBs with or without topology, and that we are able to directly infer all the salient properties of the STVB, we progress to studying their evolution through complex channels. So far, such studies have remained largely theoretical due to the lack of a direct observation toolkit. Here we show direct measurement of experimentally generated STVBs, excited by a microwave antenna source at centered at $28$ GHz and mapped by near-field probe scanning (measurement procedure described in the Supplementary Information). Since we have measured full field information, we can experimentally extract the salient features at propagation distances of $z = 0$ (near-field) and $z = 3z_R$ (far-field), where $z_R = \pi w_0^2/\lambda$ is the Rayleigh range for a beam waist along $x$-axis of $w_0$ and wavelength $\lambda = 1.07$ cm.  Figure ~\ref{fig:free_space}\textbf{a} shows the measured intensity and polarization distributions at propagation distance of $z=0$ for the non-topological (dielectric MS) and topological STVBs (metallic MS). Here we illustrate only on the $y=0$ plane as before since the same general effects occur for other $y$-planes. Note that the transverse coordinate is normalized by the spatial beam waist at each position $w_x(z)$ since $w_x$ typically increases rapidly with propagation in such a diffraction dominated channel as free space. For both STVBs, we see the measured polarisation ellipses vary across the beam profile with the intensity and polarisation both distorted upon propagation to $3z_R$. The corresponding polarization-resolved field components are shown in Figure~\ref{fig:free_space}\textbf{b}. Due to the inherent imbalance between diffraction and dispersion \cite{hancock2019free,hancock2021mode}, we see the initial doughnut structure for t-OAM beams with $|\ell|=1$ gradually decomposes into lobes with opposite spatiotemporal tilts. We note that since the V-polarized component of the topological STVB is dominated by the unmodulated zeroth-order contribution and so, in theory, should not split into multiple lobes. However the apparent lobe splitting  in Figure~\ref{fig:free_space}\textbf{b} results from the presence of additional t-OAM components. This evolution of the component modes upon propagation leads to significant crosstalk when one sends and receives in the original t-OAM basis, as evidenced by the crosstalk matrices as shown in Figure~\ref{fig:free_space}\textbf{c}. This shows the 1D modal decomposition in the t-OAM basis for the components of the STVB beam when projected onto the   ``original basis'' which consists of the ideal t-OAM modes at the $y=0$ plane. The $z = 0$ matrix clearly shows larger intensities (lighter blue blocks) for  the component modes of the two STVBs with high fidelity, with the remaining modal crosstalk due to to imperfections in the metasurface. At $z=3z_R$ the modal crosstalk increases with modal power spreading to adjacent t-OAM modes as propagation distance increases with decreased fidelity. This is because these spatiotemporal Laguerre-Gaussian spatiotemporal vortices are not eigenfunctions of the free-space paraxial wave equation. They therefore do not propagate self-similarly in free space. Interestingly, the evolved wave packets maintain parity, with even-order $\ell$ basis states primarily coupled with other even-order basis states, while odd-order wave packets ($l=\pm1$) couple with odd ones. This behavior is rooted in the spatiotemporal parity of these wave packets and the even-symmetric nature of the free-space transfer function. Since the overlap integral between functions of opposite parity vanishes under an even-parity perturbation, the $\ell=\pm 1$ components generated by our metasurface, which are both odd-symmetric inevitably exchange energy. However, because additional t-OAM components are present in the experimentally measured STVBs, meaning both odd- and even-order t-OAM contributions are non-zero within a single polarization component as per Figure~\ref{fig:free_space}\textbf{c}, the odd–even parity selectivity is not very pronounced, see SI for further details. Nevertheless, this crosstalk from Figure ~\ref{fig:free_space}\textbf{c} can be mitigated by sending in the original t-OAM basis but measuring in a new basis of distortion-adapted basis states, as shown by the block titled ``unscrambled basis'' in Figure~\ref{fig:free_space}\textbf{c}. This allows us to retrieve the modal crosstalk $z=0$ plane even at $z=3z_R$  by dynamically updating the measurement basis to match the evolved STVB as shown by the closely matching crosstalk for the original basis at $z=0$ and the unscrambled basis at $z=3z_R$.  This propagation-adapted basis completely removes the projection mismatch induced by free-space propagation, because the field is always projected onto the corresponding evolved basis states (see SI). This is the space-time equivalent of singular value decomposition, common in the spatial structured light domain, and now performed here for the first time.  It is made possible by the ability to measure the full field information directly.

\noindent \textbf{Non-separability of space-time waves in free-space.} In addition to monitoring the distortion of these underlying DoFs, we also leverage on their non-separable relationships. STVBs have recently been studied in the context of their non-separability \cite{huang2024spatiotemporal}, a form of classical entanglement \cite{spreeuw1998classical} that has proved insightful in blurring the classical-quantum divide \cite{nape2022revealing,ndagano2017characterizing}. We translate the theory for non-separabilities in spatial modes \cite{selyem2019basis} to the space-time domain, allowing us to monitor the dynamics in any channel for any form of STVB (see SI). This allows us to confirm that, because free-space propagation acts as a unitary channel, the classical entanglement quantified by non-separability remains invariant when evaluated in this ``unscrambled'' propagation-adapted detection basis, whereas evaluation in the fixed initial t-OAM basis shows an apparent decay caused by basis mismatch and projection leakage. We applied this to the free-space channel for the non-topological beam calculated from the measured MS transfer functions as an example at various propagation distances up to $10z_R$ as shown in Figure ~\ref{fig:free_space}\textbf{d} (although the same applies for the topological STVB). The blue dotted line shows that the non-separability as measured by a vector quality factor (VQF) - the classical equivalent to quantum concurrence, is maintained when measuring in this unscrambled basis, yielding a consistently high value of 0.905 (ideal value of 1), demonstrating that the intrinsic non-separability remains intact despite the redistribution of energy among specific DoFs.  Here VQF $ = 0$ corresponds to a separable field, whereas VQF $ = 1$ corresponds to a maximally non-separable field in the chosen polarization-spatiotemporal basis.  This decay is thus not a loss of intrinsic polarization-spatiotemporal non-separability, but results from projection mismatch: during free-space propagation, the spatiotemporal Laguerre-Gaussian basis states deform and their energy is redistributed into other basis states when projected onto the fixed initial basis (further details see SI). This invariant non-separability can also be retrieved in a basis-independent manner from the global Stokes parameters as shown by the black plot in Figure~\ref{fig:free_space}\textbf{d} (see SI).  This allows us to confirm that the classical entanglement as measured by non-separability is indeed invariant in free-space when measured in a basis that maintains unitarity, yet decays in a basis (OAM) that does not.

\begin{figure*}[ht!]
	\includegraphics[width=0.89\linewidth]{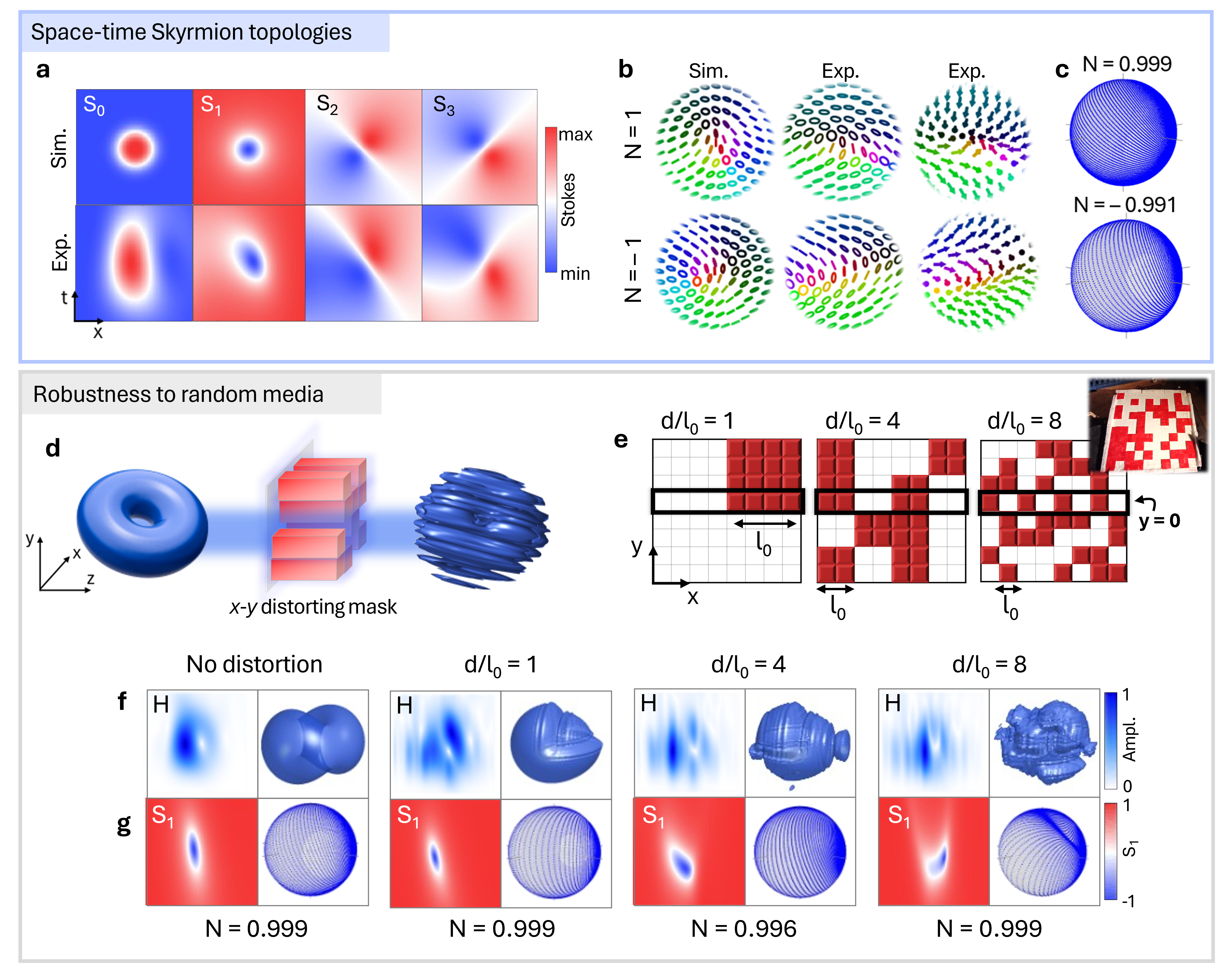}
     \caption{\textbf{Topological STVBs in random media.} \textbf{a} Simulated and measured locally renormalised Stokes parameters, $S_i$ in the $y=0$ plane for the topological STVB where $\ell_1=1$ and $\ell_2=0$. \textbf{b} Simulated and measured polarisation textures, Stokes vector textures. \textbf{c} Measured Poincar\'{e} sphere coverage for $N=\pm1$ with numbers showing the experimentally measured skyrmion number. \textbf{d} Schematic of robustness experiment: STVB passes through a spatially perturbing medium with distortions in the $x-y$ plane with measurement plane at $z_R/80$. Red blocks correspond to pillars of chosen height and material to achieve the desired absorption and phase response. Simulated H polarisation intensity iso-surfaces inferred from measured mask transmission functions and tilted for clearer visualisation. \textbf{e} Random phase and amplitude distorting masks used. Increasing distortion strengths $d/l_0 = 1,4,8$ shown where $d$ is the beam diameter and $l_0$ is the mask's correlation length in $x-y$ plane. Red blocks indicate distorting pillar regions for phase and amplitude mask. The highlighted black block indices the distorting mask at the measured $y=0$ plane.  Inset shows a photograph of one random realisation of a real constructed phase mask. \textbf{f} Measured distorted horizontal amplitude in the $y=0$ plane and  full inferred intensity iso-surfaces for various distortion strengths. \textbf{g} $S_1$ profiles and measured Poincar\'{e} sphere coverage. Measured Skyrmion number, $N$, indicated below. All data processed identically using the same interpolation and normalisation procedure. }
	\label{fig:phase_masks}
\end{figure*}

\vspace{0.2cm}
\noindent \textbf{Topological STVBs in random media.} We are now in a position to study the dynamics of STVBs in arbitrary complex media, the first to do so. Here we focus on the topological STVB and analyse this as a spatiotemporal skyrmions. We begin by generating two such topological STVBs with topological skyrmion numbers $N = \pm 1$ by using the metallic MS to generate STVBs of the form $|\pm1\rangle_{xt} |H\rangle$ +  $|0\rangle_{xt} |V\rangle$ respectively. These complementary $N=\pm1$ functionalities are realised using only the metallic MS by exploiting its directional asymmetry and operating the metasurface in the x-reversed orientation to produce both $\ell=\pm1$ t-OAM on the horizontal polarisation while the vertical polarisation remains Gaussian. Figure~\ref{fig:phase_masks}\textbf{a} shows the simulated and experimentally measured Stokes profiles, $S_i$, at the $y=0$ plane for the $N=1$ spatiotemporal skyrmion. Here we apply a standard local renormalisation of the Stokes profiles and additionally apply an established Gaussian interpolation filter which we does not change the underlying topology \cite{Peters2026extracting} with the same procedure applied for all datasets. The $S_1$ profile in Figure~\ref{fig:phase_masks}\textbf{a} shows the positive horizontally polarized t-OAM contribution (blue) around centre and with the unmodulated vertical Gaussian component dominant near the centre (red). The experimental data closely resembles the simulation where the relative phase measured between the polarisation components has been accounted for in the simulation. Since the STVB is y-invariant as mentioned above, we henceforth present analysis only at this $y=0$ plane as before.  The other Stokes parameters also closely match between simulation and experiment. Figure~\ref{fig:phase_masks}\textbf{b} shows the state of polarisation with the polarisation ellipses which vary from vertically polarised at the centre and approach horizontally polarised toward the edge of the beam matching that observed in $S_1$ from Figure~\ref{fig:phase_masks}\textbf{b}. This matches in experiment and simulation. The Stokes vector textures constructed from the Stokes vector $\textbf{S}(x,t)=[ S_1,  S_2,  S_3]^T$ are also indicated.  These locally renormalised Stokes parameters are used to compute spatio-temporal skyrmion number via a commonly used line integral approach (see SI). Figure~\ref{fig:phase_masks}\textbf{c} shows the experimentally measured full Poincar\'{e} sphere coverage confirming the presence of all the polarisation states once over and thus the non-trivial topological nature of these states for both $N=\pm1$ functionalities. The extracted skyrmion number closely matches the target wrapping number of $N=\pm1$ thus confirming the initial topology.  We note that although our metallic MS conveniently allows us to generate both $N=\pm1$, similar responses could alternatively be achieved with distinct metasurface designs for the different skyrmion numbers obtained through suitable parameter optimisation. 

Although the pristine measurements in Figure ~\ref{fig:phase_masks}\textbf{c} confirm the successful generation of the desired non-trivial spatio-temporal skyrmion topology, this does not automatically guarantee topological robustness through complex channels. While the robustness of many spatial Stokes skyrmions through select channels has been recently been demonstrated \cite{Guo2026topological, Wang2024topological}, no tests have yet been performed to see if similar robustness persists for space-time skyrmion topologies for channels beyond free-space \cite{Teng2025construction}. To test this, we pass the pristine $N=1$ topological beam through a random spatial distorting channel, apply the same procedure as in the pristine case and measure the distorted output STVB beam as shown schematically in Figure~\ref{fig:phase_masks}\textbf{d}. The spatial distorting channel is characterized by a static binary random phase and amplitude distorting mask applied in the $x-y$ plane. Phase-only masks were also tested (see SI) with similar results. These distorting masks was constructed from pillars of different heights as shown in Figure~\ref{fig:phase_masks}\textbf{d} with the STVB measurement made after a short subsequent propagation distance after the mask to the measurement plane of $z_R/80$ which allows for detector clearance as conceptually shown by the distorted output STVB in Figure ~\ref{fig:phase_masks}\textbf{d}. Three amplitude and phase distorting mask configurations were tested which allow us to examine, for the first time, the robustness of these space-time topological states under increasing levels of disorder. Figure ~\ref{fig:phase_masks}\textbf{e} shows the spatial distributions of the three different distortion strength masks with the inset showing one example of a real constructed mask.  A time lapsed video of the mask assembly is shown in the Supplementary Information.  The spatial mask's distortion strength is given by $d/l_0$, where $l_0$ represents the transverse correlation length over which the mask's action is approximately constant with smaller $l_0$ block sizes corresponding to more rapidly fluctuating mask profiles as shown in Figure~\ref{fig:phase_masks}\textbf{e}. The distortion strength is scaled relative to the beam diameter, $d$, in the $x$-direction so that larger $d/l_0$ corresponds to stronger distortion strengths. The measured output horizontal polarisation amplitude profile at the $y=0$ plane  after each mask is shown in Figure ~\ref{fig:phase_masks}\textbf{f} alongside the effect on the inferred full intensity iso-surface profile. For the case of no distortion, the characteristic donut-shaped H amplitude profile is clearly visible. As the distortion strength $d/l_0$ increases, the mask's regions of constant phase and amplitude decrease and we see increasing degradation of the beam's amplitude structure due to the mask's perturbation. Despite this degradation, the locally renormalised Stokes parameter $S_1$ shown in Figure~\ref{fig:phase_masks}\textbf{g} remains invariant. This is because the applied random mask acts identically on both orthogonal polarisation components of the STVB and so does not affect the Stokes parameters. The corresponding measured Poincar\'{e} sphere coverage and the computed topological numbers from the Stokes parameters thus also remain near $N=1$ for all distortion strengths as indicated and also shown by the complete coverage of the Poincar\'{e} sphere for $d/l_0=1,4 $ and $8$ in Figure~\ref{fig:phase_masks}\textbf{g}.   In all cases investigated here, the topology is preserved, as evidenced by the invariant skyrmion number $N$, despite substantial distortion of the underlying field distributions. We note that this robustness is demonstrated only for the class and range of spatial perturbations considered here. For sufficiently fine-scale or strongly discontinuous perturbations enhanced high-spatial-frequency scattering may compromise the continuous polarization mapping required for reliable reconstruction of the skyrmion number, however this regime is beyond the scope of the present work.

\vspace{0,2cm}
\section*{Discussion and conclusion}
Space-time wavepackets are attracting considerable interest, bridging the gap between structured light and ultrafast optics. The field has opened up exciting new forms of light, but the study of such STVBs in complex media is still emerging hindered by the difficulty of direct observation \cite{Teng2025construction}. Our work leverages on topology and non-separability of spatial light to generalise the measurement and analysis tools to the space-time domain. This parallel may open new paths to the preservation of information in STVBs mirroring advances in spatially structured light, e.g., for error-free communication using non-separability \cite{singh2023robust,cai2026free} without any knowledge of the distorting medium. We have shown that if knowledge of the medium exists, then a new space-time basis can be found that returns cross-talk free communication and that a basis independent approach can also be used for these space-time beams.  Both these leverage on the unitary and one-sided nature of the channel for space-time waves.  
For more general channels, such as our exemplary case of random media, topology is naturally invariant with no knowledge of the channel required, but this still necessitates a topological detector to be exploited, devices that may be on the horizon for spatial modes and can be extended to space-time too \cite{bezuidenhout2025deep}. 

In conclusion, we have demonstrated a simple and compact solution for the creation of topological or non-topological STVBs in a single element by symmetry breaking metasurfaces, opening an exciting path for their on-demand creation and control. Our realisation in the microwave allows us to directly observe their salient properties, which we use to present the first studies of spatiotemporal light through complex channels, showing how distortion can be overcome through appropriate measurement, and uncovering the invariances inherent in such classically entangled and topological space-time light structures. The extension to the microwave is not an inconvenience: it is the core of modern day communication systems and integral to long distance sensing.  As such, our work opens exciting prospects for fast-tracking the deployment of such using space-time structure in these real-world applications.

%%%%%%%%%%%%%%%%%%%%%%%%%%%%%%%%%%%%%%%%%%%%%%%%%%
%%% Additional info
%%%%%%%%%%%%%%%%%%%%%%%%%%%%%%%%%%%%%%%%%%%%%%%%%%
%%%%%%
\section*{Acknowledgments}
This work is sponsored by the Joint Funds of the National Natural Science Foundation of China (U24A20313), National Natural Science Foundation of China (62475234), Natural Sci
ence Foundation of Zhejiang Province LDT23F05014F05, National Key Research and Development Program of China (2024YFA1012600). A.F. and K.E. thank the CSIR Rental Pool, SA QuTI and the Oppenheimer Memorial Trust for funding.

%%%%%
\section*{Author contributions}
R.Z. built the experiment and derived the theory. R.Z and K.E. performed the numerical simulations, experiments and data analysis. All authors contributed to the writing of the manuscript. Y.M. and A.F. conceived the idea and supervised the project.

%%%%%
\section*{Competing Interests}
The authors declare no competing interests.

%%%%%%
\section*{Materials availability}
\noindent Code, data and materials are available upon reasonable request from the corresponding author.

\clearpage
%\appendix

\setcounter{section}{0}
\setcounter{figure}{0}
\setcounter{table}{0}
\setcounter{equation}{0}
\setcounter{footnote}{0}
\renewcommand{\thesection}{S\arabic{section}}
\renewcommand{\thefigure}{S\arabic{figure}}
\renewcommand{\thetable}{S\arabic{table}}
\renewcommand{\theequation}{S\arabic{equation}}

\section*{Supplementary Information: Topological space-time waves in complex channels}

\renewcommand{\theequation}{S.\arabic{equation}}

\section{Design and characterization of the asymmetric birefringent dielectric metasurface}
The generation of the target spatiotemporal vector beam (STVB) is achieved through a spectral transfer-function approach, extending the principles of paraxial beam shaping into the time domain. The basic idea is to decompose an incident spatiotemporal Gaussian pulse into its plane-wave spectral components and to engineer the metasurface response such that different polarization components acquire different spatiotemporal spectral modulations. In the present dielectric metasurface design, the x-polarized channel transforms the incident Gaussian spectrum into a first-order temporal Hermite–Gaussian component, while the y-polarized channel transforms it into a first-order spatial Hermite–Gaussian component along the x direction. Their coherent superposition gives rise to a spatiotemporal vector wave packet with polarization-dependent transverse orbital angular momentum.\par
From a wave-optics perspective, an arbitrary incident spatiotemporal wave packet can be decomposed into a superposition of monochromatic plane waves by a Fourier transform. In the spatiotemporal spectral domain, we denote the spectral coordinate as $(k_x,k_y,\Omega)$ where $k_x$ and $k_y$ are the transverse wavevector components, and $\Omega=\omega-\omega_0$ is the sideband angular frequency with respect to the carrier frequency $\omega_0$. The incident and output field can therefore be written as
\begin{equation}
       \begin{aligned}
              \tilde{\bm{E}}_{\rm in(out)}(k_x,k_y,\Omega)=\iiint & \bm{E}_{\rm in(out)}(x,y,t) \times \\ & e^{-i(k_x x+k_y y-\Omega t)}\,
       \mathrm{d}x\mathrm{d}y\mathrm{d}t.
       \end{aligned}
        \label{eq:st_FFT}
\end{equation}

\par
Since the metasurface period is chosen to be subwavelength with respect to the central operating wavelength, all higher-order diffraction channels are evanescent. Consequently, the device operates in the zeroth-order transmission regime. Under this condition, the metasurface can be modeled as a linear spectral transfer system, where each plane-wave component is multiplied by a complex transmission coefficient. In the linear polarization basis $(\bm{e}_x,\bm{e}_y)$, the metasurface is designed to exhibit negligible direct polarization conversion between the $x$- and $y$-polarized channels. Its Jones-type transfer matrix can therefore be expressed as
\begin{equation}
       H_{xy}(\bm{q})=
        \begin{pmatrix}
        H_x(\bm{q}) & 0 \\
        0 & H_y(\bm{q})
\end{pmatrix}
        \label{eq:TF_matrix}
\end{equation}
where $\bm{q}=(k_x,k_y,\Omega)$. Here, $H_x(\bm{q})$ and $H_y(\bm{q})$ are the complex transfer functions for the $x$- and $y$-polarized channels, respectively. Therefore 
\begin{equation}
\tilde{\bm{E}}_{\mathrm{out}} = H_{xy}(\bm{q})\tilde{\bm{E}}_{\mathrm{in}}(\bm{q}).
\end{equation}
For the selected incident Gaussian pulse, the input polarization is taken as a tilted linear polarization state,
\begin{equation}
      \tilde{\bm{E}}_{\mathrm{in}}(\bm{q})=\mathrm{GS}(\bm{q})\begin{pmatrix}a\\b\end{pmatrix}
        \label{eq:input_pol}.
\end{equation}
where $\mathrm{GS}(\bm{q})$ is the scalar spatiotemporal Gaussian spectrum. For a purely linear incident polarization, $a$ and $b$ can be chosen as real coefficients satisfying $a^2+b^2=1$. 
For the output field, a general pair of orthonormal left-handed elliptical polarization (LEP) and right-handed elliptical polarization (REP) basis vectors can be written as
\begin{equation}
    \bm{e}_L = \begin{pmatrix}\cos(\theta) \\
e^{i\delta}\sin(\theta)\end{pmatrix} \quad \text{and}
\quad
\bm{e}_R = \begin{pmatrix}\sin(\theta) \\
-e^{i\delta}\cos(\theta)\end{pmatrix}.
\end{equation}

The output components analyzed in the LEP and REP channels are obtained by projecting the transmitted field onto the corresponding polarization analyzers
\begin{equation}
    \tilde{\bm{E}}_{\mathrm{L}}(\bm{q})=\bm{e}_{\mathrm{L}}^\dagger \tilde{\bm{E}}_{\mathrm{out}}(\bm{q}) \quad \text{and}
\quad \tilde{\bm{E}}_{\mathrm{R}}(\bm{q})=\bm{e}_{\mathrm{R}}^\dagger \tilde{\bm{E}}_{\mathrm{out}}(\bm{q}).
\end{equation}

Substituting the transmitted field into the above equations gives
\begin{equation}
\begin{split}
\tilde{E}_{\mathrm{L}}(\bm{q}) &= \mathrm{GS}(\bm{q})\bigl[a\cos(\theta)H_x(\bm{q})+be^{-i\delta}\sin(\theta)H_y(\bm{q})\bigr], \\
\tilde{E}_{\mathrm{R}}(\bm{q}) &= \mathrm{GS}(\bm{q})\bigl[a\sin(\theta)H_x(\bm{q})-be^{-i\delta}\cos(\theta)H_y(\bm{q})\bigr].
\end{split}
\label{eq:EL_ER}
\end{equation}
We therefore define the LEP- and REP-projected transfer spectra shown in Figure 1 of the main text for the dielectric metasurface as:
\begin{equation}
\begin{split}
T_\mathrm{L}^{\mathrm{proj.}}(\bm{q}) &= a\cos(\theta)H_x(\bm{q})+be^{-i\delta}\sin(\theta)H_y(\bm{q}) \\
T_\mathrm{R}^{\mathrm{proj.}}(\bm{q}) &= a\sin(\theta)H_x(\bm{q})-be^{-i\delta}\cos(\theta)H_y(\bm{q})
\end{split}
\label{eq:TL_TR}
\end{equation}
It should be emphasized that $T_\mathrm{L}^{\mathrm{proj.}}$ and $T_\mathrm{R}^{\mathrm{proj.}}$ are not intrinsic Jones matrix elements of the metasurface in the LEP/REP basis. Instead, they are scalar spectral responses obtained after specifying the incident polarization state and projecting the transmitted field onto the LEP and REP analyzers. They characterize the LEP and REP components contained in the output field under the selected incident polarization.\par
In the ideal case, the two linear polarization channels are designed to satisfy $\mathcal{F}^{-1}\{H_x(\bm{q})\mathrm{GS}(\bm{q})\}\propto \mathrm{st}\mathrm{HG_{0,1}}$ and $\mathcal{F}^{-1}\{H_y(\bm{q})\mathrm{GS}(\bm{q})\}\propto \mathrm{stHG_{1,0}}$ where stHG indicates a spatio-temporal Hermite-Gaussian mode.  For a diagonal linearly polarized input $a=b=1/\sqrt{2}$, and for circular polarization analyzers, $\theta=\pi/4$, $\delta = \pi/2$. Eq. \eqref{eq:EL_ER} gives 
\begin{equation}
    E_\mathrm{LCP} ( x,t)\propto\mathrm{stHG}_{0,1}-i\mathrm{stHG}_{1,0}
\end{equation}
and
\begin{equation}
    E_\mathrm{RCP} (x,t)\propto\mathrm{stHG}_{0,1}+i\mathrm{stHG}_{1,0}.
\end{equation}
 According to the standard relation between first-order Hermite–Gaussian and Laguerre–Gaussian modes, the above two components correspond to first-order spatiotemporal Laguerre–Gaussian wave packets
 with opposite topological charges, up to an unimportant global phase. Therefore, in the ideal limit, the diagonally polarized Gaussian input is converted into a vectorial superposition of LCP and RCP components carrying opposite transverse orbital angular momentum.\par
In the actual metasurface, however, the simulated and measured transfer functions inevitably deviate from the ideal differentiation operators. Consequently, the incident linear polarization state and the output LEP/REP analyzer basis are slightly adjusted from the ideal diagonal/circular basis to maximize the modal purity and to compensate for the residual amplitude and phase mismatch between the two polarization channels. For the projected transfer spectra shown in the main text the simulated transfer functions were evaluated using $a=0.991$, $b=0.133$, $\theta = 0.15\pi$, $\delta = 0.52\pi$ whereas the experimentally measured transfer functions were projected using a different set of parameters, $a=0.983$, $b=0.183$, $\theta = 0.17\pi$, $\delta = 0.28\pi$. \par
\subsection{Engineering polarization dependent spectral responses for the dielectric MS}
\begin{figure*}[hpt!]
	\includegraphics[width=\linewidth]{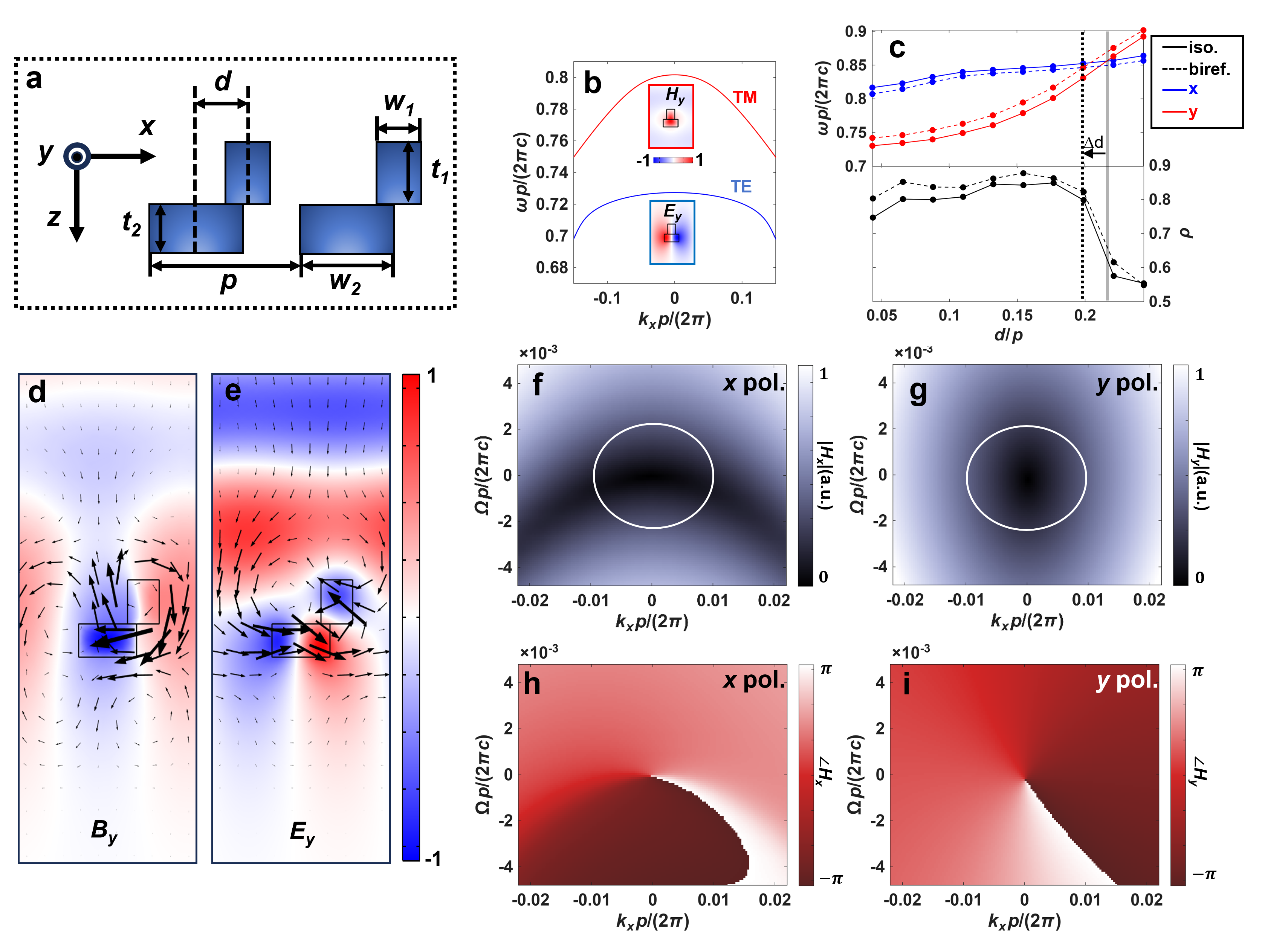}
	\caption{\textbf{Dielectric metasurface design and simulation results.} \textbf{a} Cross section schematic of sub-wavelength structure with geometric parameters indicated. \textbf{b} Band structure of the one-dimensional grating when the asymmetry factor $d=0$ and $\epsilon_x=\epsilon_y=9.8$. Blue curves correspond to TE modes, red curves to TM modes. Below the band curves, the $y$-components of the electric and magnetic fields of the TE antisymmetric and TM symmetric eigenmodes at the $\Gamma$ point are plotted. \textbf{c} Upper panel: transmission zeros for $x$ polarization (blue) and $y$ polarization (red) as a function of $d$. Lower panel: average cosine similarity $\rho$ between the two polarization transfer functions and the ideal transfer function versus $d$. Solid and dashed lines correspond to $\epsilon_x=\epsilon_y=9.8$ and $\epsilon_x=11.5$, $\epsilon_y=9.4$, respectively. \textbf{d} $y$ component of the magnetic field $B_y$ and \textbf{e} electric field $E_y$ excited by a normally incident plane wave. Black arrows indicate the Poynting vector. \textbf{f} Amplitude of the transfer function for the asymmetric grating obtained from simulation under $x$ polarization $|H_x|$ and \textbf{g} for $y$ polarization $|H_y|$. \textbf{h} Phase of the transfer function for the asymmetric grating obtained from simulation under $x$ polarization and for $y$ polarization in \textbf{i}.}
	\label{fig:S1}
\end{figure*}

To implement the distinct spectral responses required for the two polarizations for the dielectric MS, we introduce a structural asymmetry into the unit cell, as illustrated in Figure~\ref{fig:S1}\textbf{a}. The operational principle relies on the precise interplay between symmetry breaking and eigenmode engineering around the $\Gamma$ point (normal incidence). We start with a symmetric unit cell (asymmetry parameter $d=0$) which supports a symmetric transverse magnetic (TM) mode and an antisymmetric transverse electric (TE) mode at the $\Gamma$ point as shown in Figure~\ref{fig:S1}\textbf{b}. The lateral offset between the upper and lower grating bars breaks the mirror symmetries of the unit cell with respect to both the $x=0$ and $z=0$ planes. This structural asymmetry serves two critical purposes tailored to our specific differentiation requirements. First, the symmetry breaking induces a necessary phase disparity between the transmission coefficients for plane waves with opposite transverse wave-vectors ($+k_x$ and $-k_x$) in the vicinity of the $\Gamma$ point. This phase behavior is essential for synthesizing the odd-symmetry transfer function required for spatial differentiation. Second, the symmetry breaking transforms the original antisymmetric TE eigenmode from a dark mode, namely a bound state in the continuum (BIC), into a leaky mode that couples to far-field plane waves, thereby producing a transmission dip in the metasurface's transfer function. Crucially, unlike typical high $Q$ quasi-BIC designs that rely on minute perturbations, we employ a large asymmetric parameter $d$. This deliberate design choice significantly lowers the $Q$ factor of the resonance, effectively moving the system out of the quasi-BIC regime. The reduced $Q$ factor broadens the resonance linewidth, allowing the transmission response near the zero point to better approximate a first-order spatial differentiation operator (for the y-polarization component) over the finite $(k_x, \Omega)$ bandwidth of the incident pulse. For $x$ polarization, this symmetry breaking does not significantly modify the far-field response of the TM mode, which is already a leaky mode. As a result, it still retains a typical transmission dip that follows the parabolic curvature of the band dispersion, thereby approximately realizing temporal differentiation within the finite bandwidth.\par

With the geometric asymmetry established, a further challenge is to align the operational frequencies of the two polarization channels. In a purely geometric design, the spatial differentiation zero for $y$-polarization and the temporal differentiation resonance for $x$-polarization typically occur at different central frequencies. While adjusting the assymetry parameter $d$ affects the resonance positions, it also alters the $Q$ factors and phase slopes. To decouple these effects and ensure frequency matching at $\omega_0$, we leverage the intrinsic birefringence of the grating material (sapphire). Figure \ref{fig:S1}\textbf{c} (top panel) maps the trajectory of the transmission zeros for both polarizations as a function of $d$. By tuning the material anisotropy ($\epsilon_x=11.5$, $\epsilon_y=9.4$), we shift the dispersion curves (dashed lines) such that the transmission zeros coincide at an better asymmetry parameter ($d=0.199p$), where $p$ is the grating period. At this operating point, the transfer functions achieve high fidelity compared to ideal differentiation operators, with an average cosine similarity of $\rho=0.822$.\par
Full-wave simulations using the Finite Element Method (FEM) confirm the efficacy of this design. The optimized device parameters are widths $w_1=0.176p$, $w_2=0.322p$, and thicknesses $t_1=0.220p$, $t_2=0.167p$. The calculated transfer functions are shown in Figure \ref{fig:S1}\textbf{f-i}. The $x$-polarized component (Figure \ref{fig:S1}\textbf{f,h}) exhibits a pronounced dip along the $k_x$ axis, implementing the temporal differentiation, while the $y$-polarized component (Figure \ref{fig:S1}\textbf{g,i}) displays a clear dip along the $\Omega$ axis, implementing the spatial differentiation. The white circles denote the $1/e^2$ contour of the incident Gaussian pulse spectrum, demonstrating that the engineered transmission zeros are perfectly aligned with the pulse energy, thereby enabling the successful generation of the desired STVB.\par

\subsection{Metallic metasurface engineering}
A related symmetry-breaking strategy was also adopted in our previously reported quasi-BIC metallic metasurface, where vertically staggered metallic elements were used to introduce the required structural asymmetry. That design, however, was intended for a different functionality as per the main text: an $x$-polarized Gaussian input was converted into a first-order spatiotemporal vortex, whereas a $y$-polarized input was almost unmodulated. The detailed design principle and experimental characterization of that metallic metasurface have been reported in our previous work \cite{zhou2025quasi}.\par

 \section{Experiment setup for measuring metasurface transfer functions}

\begin{figure*}[hpt!]
	\includegraphics[width=0.95\linewidth]{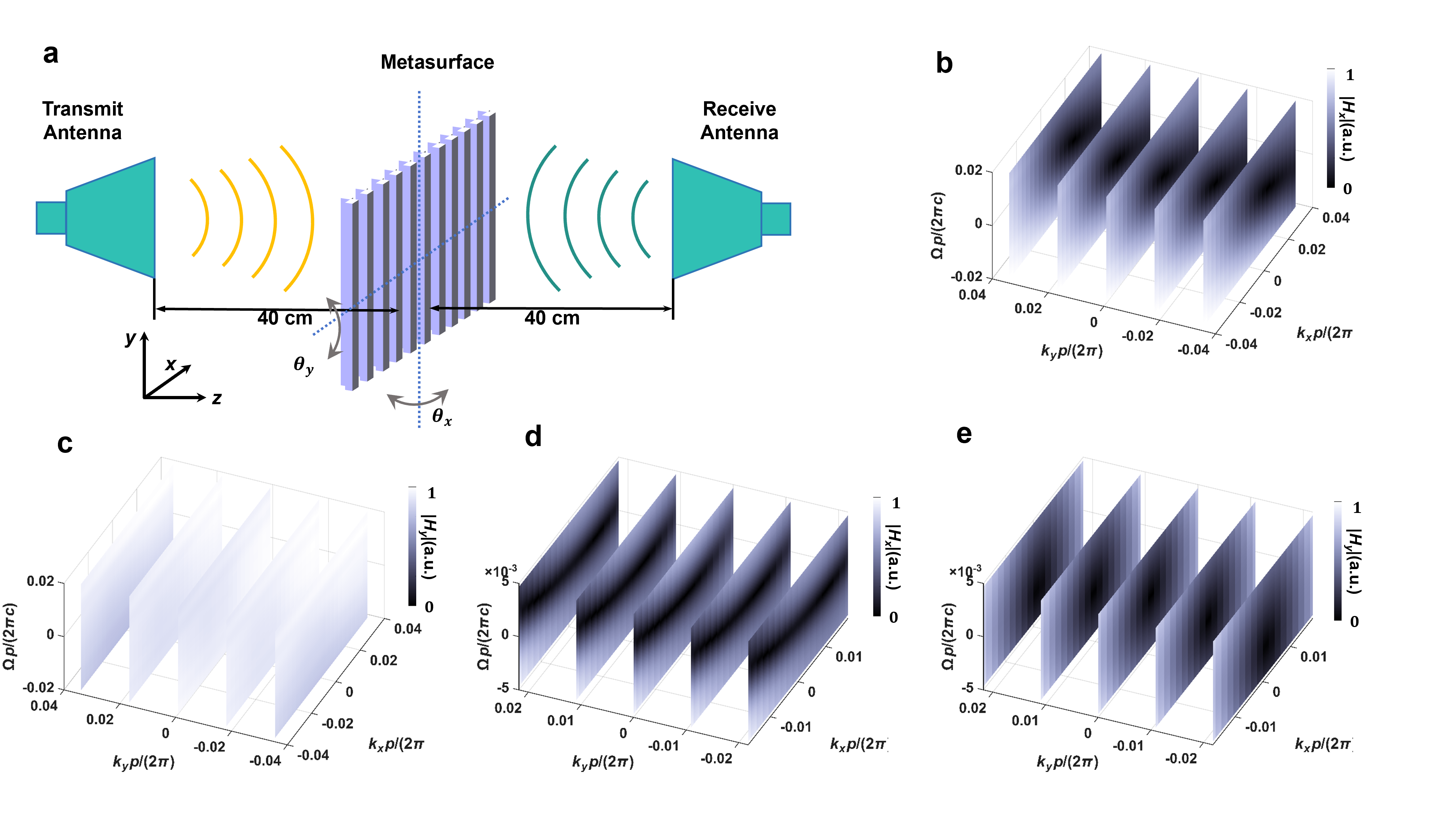}
	\caption{\textbf{Experimental measurement method and results of the transfer functions.} \textbf{a} Schematic of the experimental setup for transfer function measurement; the sample is fixed on a three‑axis rotation stage. \textbf{b} and \textbf{c} show the experimentally measured $|H_{x}\left( k_{x},k_{y},\Omega \right)|$ and $|H_{y}\left( k_{x},k_{y},\Omega \right)|$ of the metallic metasurface, respectively. \textbf{d} and \textbf{e} show the experimentally measured $|H_{x}\left( k_{x},k_{y},\Omega \right)|$ and $|H_{y}\left( k_{x},k_{y},\Omega \right)|$ of the dielectric metasurface, respectively.}
	\label{fig:S2}
\end{figure*}
 
The three-dimensional transfer functions of the metasurfaces were experimentally characterized using the setup shown in Figure \ref{fig:S2}\textbf{a}. A transmitting antenna and a receiving antenna were placed on the two sides of the metasurface, with a distance of approximately 40 cm between each antenna and the sample. The sample was mounted on a high-precision three-axis manual rotation stage. This far-field configuration ensures that the incident wavefront impinging on the metasurface can be approximated as a plane wave with a well-defined wavevector. By rotating the sample around the $y$ and $x$ axes (shown as $\theta_x$ and $\theta_y$ in Figure \ref{fig:S2}\textbf{a}), the incident plane-wave components with different transverse wavevectors $(k_x,k_y)$ were selected. For each angular configuration, the complex transmission spectrum was measured using a vector network analyzer (VNA), giving the frequency-dependent response as a function of $\Omega$. The transfer functions for the $x$- and $y$-polarized channels were obtained by simultaneously rotating the linearly polarized transmitting and receiving antennas. The measured angular and spectral responses were then converted into the three-dimensional transfer functions $H_x(k_x,k_y,\Omega)$ and $H_y(k_x,k_y,\Omega)$, which were subsequently used for the reconstruction of the 4D wave packets. \par
Figures \ref{fig:S2}\textbf{b} and \textbf{c} show the measured amplitudes $|H_x(k_x,k_y,\Omega)|$ and $|H_y(k_x,k_y,\Omega)|$ of the metallic metasurface, respectively. The $x$-polarized channel exhibits a pronounced spatiotemporal spectral modulation associated with the generation of a first-order spatiotemporal vortex with transverse OAM as shown in Figure \ref{fig:S2}\textbf{b}, whereas the $y$-polarized channel remains nearly uniform, indicating weak modulation for $y$-polarized incidence as per Figure \textbf{c}. Figures \ref{fig:S2}\textbf{d} and \textbf{e} show the measured amplitudes $|H_x(k_x,k_y,\Omega)|$ and $|H_y(k_x,k_y,\Omega)|$ of the birefringent dielectric metasurface. The $x$-polarized channel shows a dip mainly along the $k_x$ direction, corresponding to temporal differentiation, while the $y$-polarized channel shows a dip mainly along the frequency-detuning direction $\Omega$, corresponding to spatial differentiation along $x$. These measured transfer functions confirm that the two metasurfaces provide the designed polarization-dependent spatiotemporal spectral responses required for the generation and reconstruction of the target STVBs. Within the relatively small $k_y$ range covered by the incident pulse spectrum, the measured transfer functions vary only weakly along $k_y$, so the reconstructed STVBs approximately retain a Gaussian intensity profile along the $y$ direction.

\section{Near-field scanning measurement of experimentally generated STVBs}
\subsection{Dielectric metasurface and non-topological STVB}
\begin{figure*}[hpt!]
	\includegraphics[width=\linewidth]{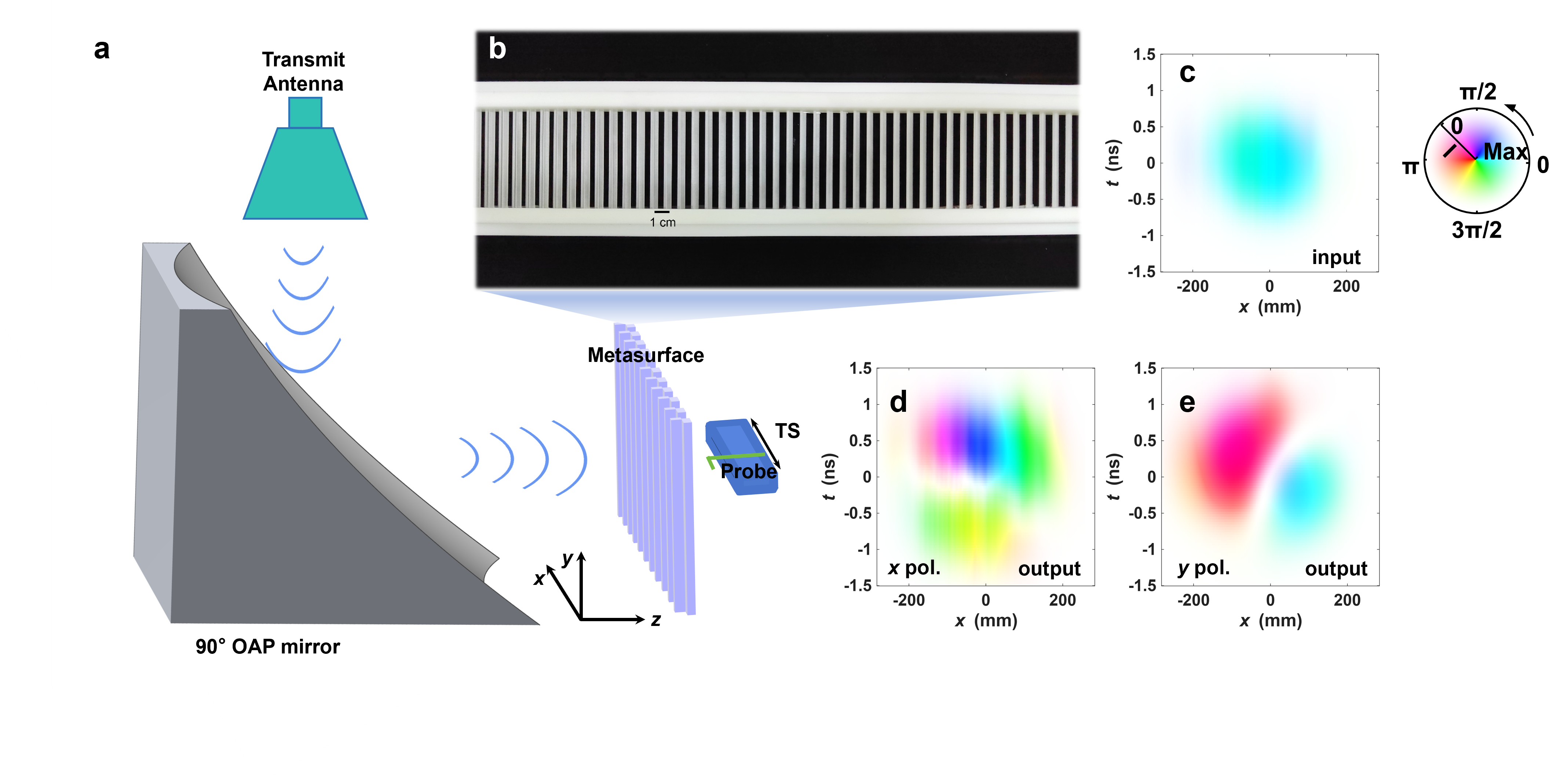}
	\caption{\textbf{Experimental setup and near-field scanning results for the generation of the non-topological STVB.} \textbf{a} Experimental setup, consisting of a transmitting antenna, a $90^\circ$ off-axis parabolic (OAP) mirror, the birefringent dielectric metasurface, a translation stage (TS), and a monopole probe. \textbf{b} Image of a local region of the metasurface sample. Grating bars separated by 1 cm. \textbf{c} Total intensity and phase distribution of the incident pulse. \textbf{d} Measured $x$-polarized component of the transmitted pulse. \textbf{e} Measured $y$-polarized component of the transmitted pulse.}
	\label{fig:S3}
\end{figure*}

To generate and characterize the non-topological STVB generated by the dielectric metasurface, we used the experimental setup shown in Figure \ref{fig:S3}\textbf{a}. A transmitting antenna, rotated in the transverse plane to set the desired incident linear polarization, emitted a linearly polarized signal. The emitted wave was expanded and collimated by a $90^\circ$ off-axis parabolic (OAP) mirror, which was fabricated by 3D printing and coated with aluminum foil on the reflecting surface, before impinging on the birefringent dielectric metasurface sample shown in Figure \ref{fig:S3}\textbf{b}. The transmitted near field was measured approximately 1 cm behind the sample using a monopole probe. Different polarization components were measured by rotating the monopole probe. The monopole probe and the transmitting antenna were connected to the two ports of a vector network analyzer (VNA). By scanning the translation stage along the $x$ direction, the complex frequency spectra at different transverse positions were obtained. \par
The measured complex spectra were processed using a frequency-domain pulse synthesis procedure. A complex spectral weight $W(f)$ was first obtained from the reference spectrum measured at $x=0$ without the metasurface sample, such that the weighted reference spectrum matched a target Gaussian spectrum. This same weight was then applied to all measured spectra at different $x$ positions, both with and without the sample. Owing to the linear time-invariant nature of the measurement system, this operation is equivalent to coherently synthesizing the response to a Gaussian incident pulse from the measured monochromatic frequency responses. The weighted spectra were then inverse Fourier transformed to obtain the $x-t$ field distributions. Figure \ref{fig:S3}\textbf{c} shows the $y=0$ plane measurement of the intensity and phase distribution of the incident pulse at the measurement position in the absence of the metasurface. Figures \ref{fig:S3}\textbf{d} and \ref{fig:S3}\textbf{e} show the measured $x$- and $y$-polarized components of the transmitted pulse with the dielectric metasurface inserted, respectively.\par

\subsection{Metallic metasurface and topological STVB}
\begin{figure*}[hpt!]
	\includegraphics[width=\linewidth]{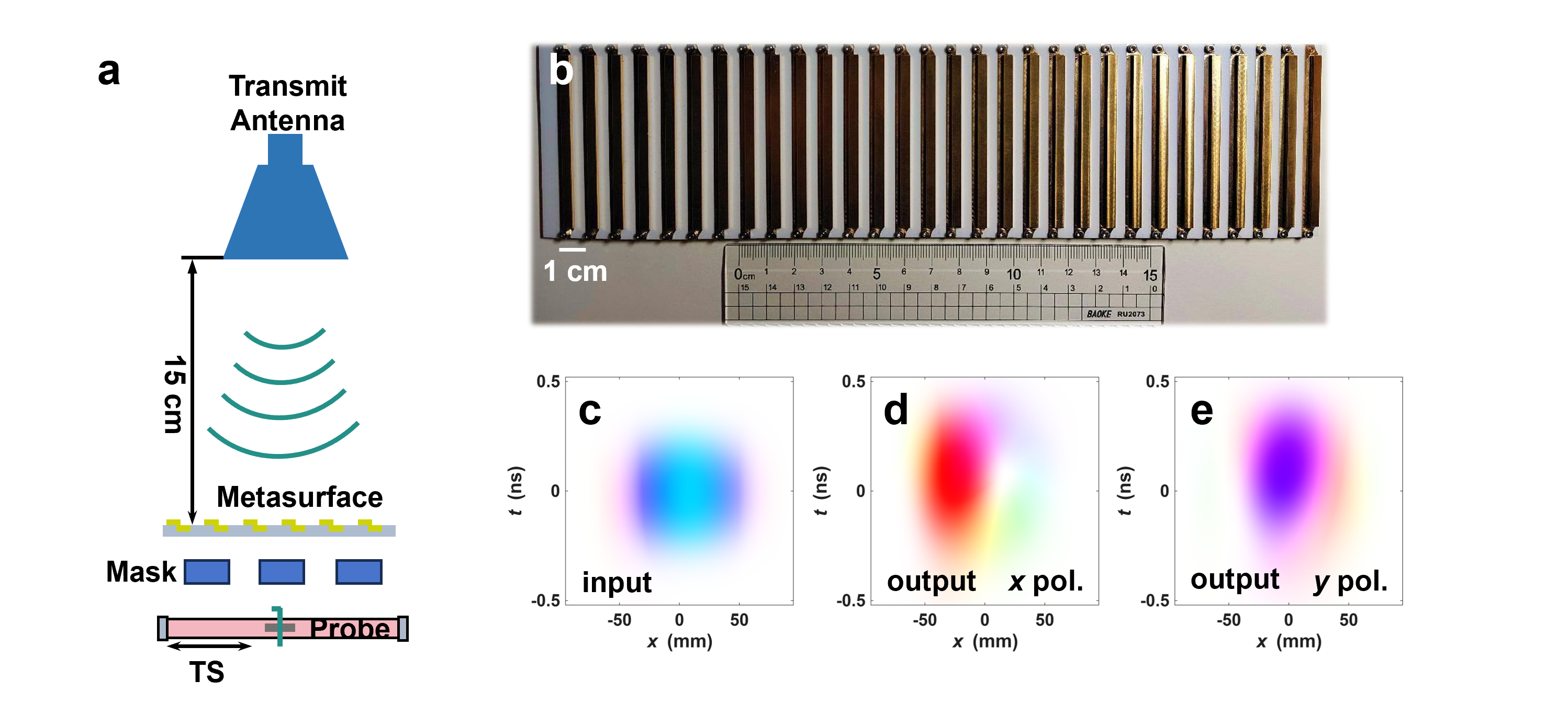}
	\caption{\textbf{Experimental setup for the generation of the topological STVB and its propagation through a complex channel.} \textbf{a} Experimental setup, consisting of a transmitting antenna, a metallic metasurface, a microwave absorbing-foam mask, a translation stage (TS), and a monopole probe. \textbf{b} Photograph of the metallic metasurface sample. \textbf{c} Total intensity and phase distribution of the incident pulse. \textbf{d, e} Intensity and phase distributions of the $x$- and $y$-polarized components of the pulse transmitted through the metasurface respectively when no distorting mask was inserted.}
	\label{fig:S4}
\end{figure*}

Figure \ref{fig:S4}\textbf{a} shows the experimental setup and representative near-field scanning results for the generation of the topological STVB and its propagation through a complex channel. The measurement procedure is the same as that used for the non-topological STVB.  Without inserting the mask, the incident pulse and the $x$- and $y$-polarized components of the pulse transmitted through the metasurface were shown in Figure \ref{fig:S4}\textbf{c}-\textbf{e}, respectively. The results measured after the mask are presented in Figure 4\textbf{f,g} of the main text. 

\subsection{Distorting masks}
\begin{figure}[hpt!]
	\includegraphics[width=\linewidth]{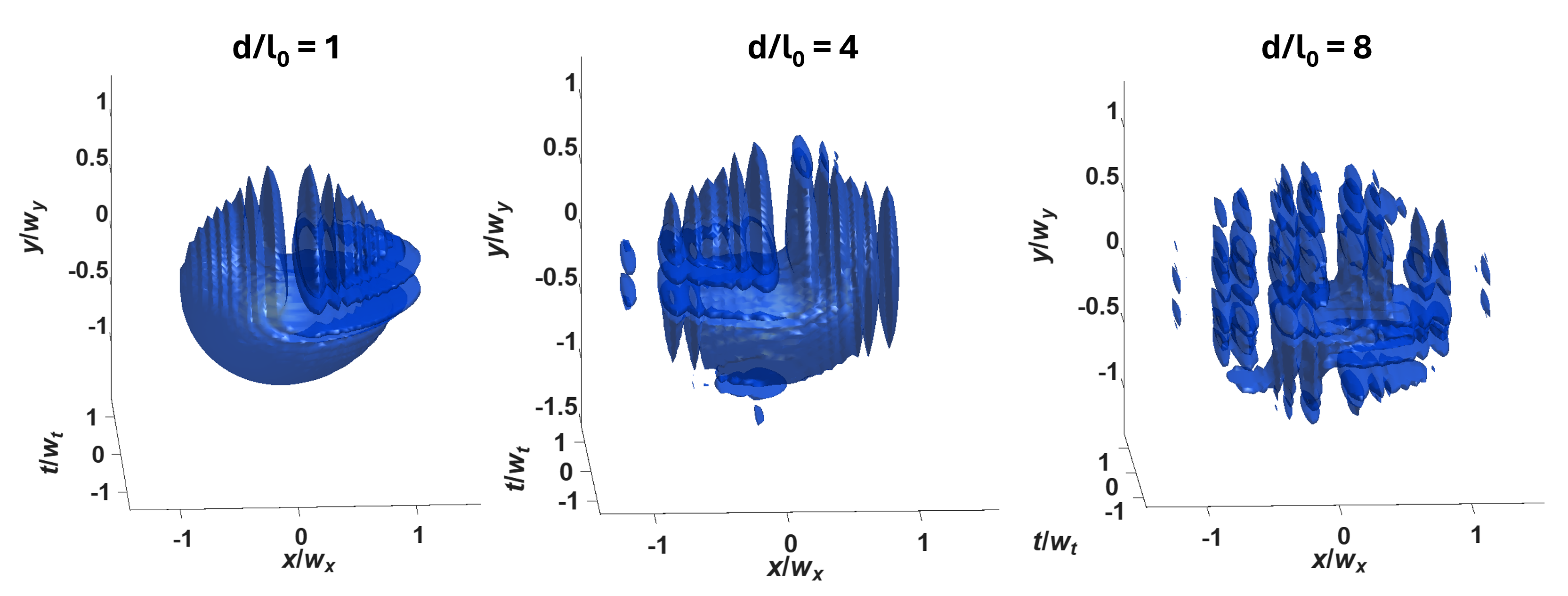}
	\caption{\textbf{Isointensity surfaces after measured phase mask transmission.} Topological STVB passed through the three random binary phase screens with distortion strengths $d/l_0$ and mask configurations shown in the main text with distorted inferred isointensity surfaces due to the diffraction effects caused by the sharp fabricated phase mask edges.}
	\label{fig:phase_screens}
\end{figure}

For the phase and amplitude distorting masks shown in the main text, a cut microwave absorbing-foam mask was placed approximately $1$ cm behind the metasurface to form the complex propagation channel. Because the material is microwave absorbing, each foam block introduces both amplitude attenuation and phase retardation, so the masks act as phase-amplitude perturbations. Specifically in each distorting ``pillar'' an approximate phase modulation of $\pi/2$ and a transmittance of approximately $16\%$ was imparted onto the beam. To construct the complex transmission channels used in the topology-robustness experiment, microwave absorbing foam was cut into block-like masks with the chosen distortion strength $d/l_0$. Three different channels were fabricated by changing the transverse size of an individual foam block, as schematically shown in Figure 4\textbf{e} of the main text. In this work, the topological STVB beam diameter was taken as $d=104$ mm, and the channel disorder was parameterized by $d/l_0$, where $l_0$ denotes the transverse size of a single foam-mask unit. The foam thickness was not perfectly uniform, which can introduce additional spatial variations in both attenuation and phase delay across the mask. Since the transverse scale of each foam unit is only several wavelengths and becomes smaller for larger $d/l_0$, the finite and asymmetric geometry of the foam mask can also introduce a certain degree of anisotropic distortion. Nevertheless, under our experimental conditions, the measured skyrmion number remains nearly unchanged, indicating that the topological structure is preserved despite these channel-induced distortions. As a result, the locally normalized $\mathrm{S}_1$ distribution of the STVB transmitted through the complex channel slightly deviates from that measured in the air channel, and the deviation becomes more pronounced as $d/l_0$ increases although the polarisation structure required to calculate the skyrmion number is not destroyed. We also produced phase-only distorting masks of various distorting strengths $d/l_0$. These were binary random phase screens introducing only $0$ or $\pi$ phase jumps fabricated using high precision 3D printing technology. The screens were constructed from photo
sensitive resin with a measured relative permittivity of $\varepsilon_r\approx2.7$. The binary phase modulation was physically implemented by printing an array of dielectric blocks with two distinct heights $h_1 = 9.2$ mm and $h_2 = 0.9$ mm. The height difference $h = 8.3$ mm was calculated to induce a phase delay of at the central operating frequency. These dielectric blocks were mounted on a low permittivity foam substrate ($\varepsilon_r\approx1$), which is transparent to electromagnetic waves, ensuring that the phase modulation arises solely from the printed structures. We fabricated three distinct screens with varying pixel sizes to emulate different scattering strengths (weak, moderate, and strong disturbance) with one image of the real sample as an example shown in the main text and in the supplementary video showing the construction of the phase mask. The complex transmission matrices of these phase screens were characterized using a near-field scanning microwave system described above. Utilizing the experimentally measured phase screen data, we reconstructed the evolution of the spatiotemporal vector wave packets through the scattering channel. Subsequently, to visualize the resultant intensity speckles and diffraction effects, the field is propagated over a short distance $z = z_R/80$ in free space as in the main text. This propagation distance is chosen to be sufficiently large to allow the phase discontinuities to convert into observable amplitude fluctuations as is clear in figure \ref{fig:phase_screens} due to diffraction around the sharp fabricated mask edges, yet sufficiently small to ensure that the intrinsic Gouy phase evolution of the underlying basis components is negligible.

\section{Theoretical Framework for Spatiotemporal Vector Quality Factor}
 Here we extend the spatial theory for non-separability \cite{nape2022revealing} to the space-time domain.  Our 4D spatiotemporal (ST) vector wave packet can be described in a composite Hilbert space $H = H_p \otimes H_{ST}$, formed by the tensor product of the polarization degree of freedom (DoF) $H_p$, and the spatiotemporal DoF $H_{ST}$. A general pure STVB state restricted to a two-dimensional spatiotemporal subspace can be written as

 \begin{eqnarray}\label{eq:pure_state}
 |\Psi\rangle &=a|e_1\rangle_\mathrm{p}|u_1\rangle_\mathrm{ST} +  b|e_1\rangle_\mathrm{p}|u_2\rangle_\mathrm{ST} +  \nonumber \\ & c|e_2\rangle_\mathrm{p}|u_1\rangle_\mathrm{ST} +  d|e_2\rangle_\mathrm{p}|u_2\rangle_\mathrm{ST} .
  \end{eqnarray}

 Here $|e_1\rangle_\mathrm{p}$ and $|e_2\rangle_\mathrm{p}$ are two orthonormal polarization basis states, while $|u_1\rangle_\mathrm{ST}$ and $|u_2\rangle_\mathrm{ST}$ are two orthonormal spatiotemporal basis states. The complex coefficients satisfy $|a|^2 + |b|^2 + |c|^2 + |d|^2 = 1$.
The non-separability between the polarization and spatiotemporal degrees of freedom is quantified by the vector quality factor (V), which is mathematically equivalent to the concurrence of a pure two-qubit state
 \begin{equation}\label{eq:basis_dependent_VQF}
V = 2|ad-bc|.
 \end{equation}
Here $V = 0$ corresponds to a separable field, whereas $V = 1$ corresponds to a maximally non-separable field in the chosen polarization-spatiotemporal basis. This quantity is basis-dependent because it requires a pre-defined pair of spatiotemporal basis states, $|u_1\rangle_\mathrm{ST}$ and $|u_2\rangle_\mathrm{ST}$.

 In the context of STVBs, the spatial degrees of freedom $(x,y)$  and the temporal degree of freedom $t$ possess distinct physical units and vastly different scales. To rigorously define the theoretical spatiotemporal basis functions and ensure they are mathematically well-behaved, it is standard practice to normalize the physical coordinates into a dimensionless domain.We introduce the normalized dimensionless coordinates $(\xi, \eta, \tau)$ scaled by the characteristic beam parameters $\xi=x/w_x$ , $\eta=y/w_y$  and $\tau=t/w_t$ , where $w_x$ and $w_y$ represent the characteristic spatial beam waist radius, and $w_t$ represents the characteristic pulse half-width (at $1/e^2$ intensity or similar definition depending on the specific HG/LG mathematical basis used). In this normalized spatiotemporal frame, the orthogonality condition for the theoretical basis states $\{ |u_n\rangle\}$ is defined as
\begin{eqnarray}
    \langle u_i | u_j \rangle =& \iiint_{-\infty}^{+\infty} u_i^*(\xi, \eta, \tau) u_j(\xi, \eta, \tau) d\xi d\eta d\tau \nonumber \\&= \delta_{ij}.
\end{eqnarray}

 To compute the basis-dependent V defined in Eq. \eqref{eq:basis_dependent_VQF}, we must extract the complex coefficients $a,b,c,d$ from the target 4D field distribution. The total 4D field $\left| \Psi \right\rangle$ is first projected onto the orthogonal polarization basis states $\left| e_{1} \right\rangle$, $\left| e_{2} \right\rangle$ (e.g., $\left| H \right\rangle$ and $\left| V \right\rangle$). This yields two scalar spatiotemporal fields, $\left| \psi_{1} \right\rangle = \left\langle e_{1} \middle| \Psi \right\rangle$ and $\left| \psi_{2} \right\rangle = \left\langle e_{2} \middle| \Psi \right\rangle$. By calculating the overlap integrals (inner products) of these scalar fields with the ST basis states, the four unnormalized complex coefficients 
 \begin{eqnarray}\label{eq:calc_coeff}
 c_{mn} =& \left\langle u_{n} \middle| \psi_{m} \right\rangle \nonumber \\=& \iiint u_{n}^{*}(\xi,\eta,\tau)\psi_{m}(\xi,\eta,\tau)\,d\xi\,d\eta\,d\tau 
 \end{eqnarray}
 can be obtained. Finally, the coefficients are normalized to satisfy the probability conservation condition, 
 \begin{eqnarray}
 a = {c_{11}}/{\sqrt{N}},\quad b = {c_{12}}/{\sqrt{N}}  \nonumber \\ 
c = {c_{21}}/{\sqrt{N}},\quad d = {c_{22}}/{\sqrt{N}}
 \end{eqnarray}
 where $N = \sum |c_{ij}|^{2}$, $(i,j \in (1,2))$. These normalized coefficients are then used to calculate Eq. \eqref{eq:basis_dependent_VQF}.

 While the definition of the basis functions $u_{n}(\xi,\eta,\tau)$ requires appropriate scaling factors ($w_{x},w_{y},w_{t}$) to match the shape of the analyzed pulse, the calculation of the vector quality factor itself proves to be invariant to the weighting of the integration measure used in Eq. \eqref{eq:calc_coeff}. A change in coordinate scaling simply introduces a constant Jacobian factor $J = w_{x}w_{y}w_{t}$ to all unnormalized coefficients $c_{ij}$. Since the final coefficients $a,b,c,d$ are derived via normalization, this common factor $J$ appears in both the numerator and the denominator, cancelling out perfectly. Consequently, the V analysis can be performed directly using raw data coordinates without explicit unit conversion, provided the theoretical basis functions are correctly scaled to match the pulse shape.

Alternatively, the basis-independent V can be determined without prior knowledge of the specific spatiotemporal basis by generalizing the Stokes tomography method to the 4D domain. Following the framework for spatial vectorial beams \cite{selyem2019basis}, we define the global Stokes parameters $\overline{S}$ for spatiotemporal waves by integrating the local Stokes densities $S_{i}(x,y,t)$ over the entire pulse volume ${\overline{S}}_{i} = \iiint S_{i}(x,y,t)\,dx\,dy\,dt$. These global parameters correspond to total energy measurements projected onto the respective polarization basis states, which are naturally acquired by standard detectors that integrate photon flux over the measurement window. V is then extracted from the length of the global Bloch vector
 \begin{equation}\label{VQF_base_independent}
 V = \sqrt{1 - \frac{\sum_{k = 1}^{3}{\overline{S}}_{k}^{2}}{{\overline{S}}_{0}^{2}}}.
 \end{equation}
 
 Crucially, this metric shares the same scale-invariance property as the basis-dependent method. Since V depends solely on the ratio of the squared Stokes parameters, any coordinate scaling factor $J$ introduced by the integration measure appears in both the numerator and the denominator and subsequently cancels out. This property renders the basis-independent vector quality factor a robust witness for spatiotemporal non-separability, capable of characterizing the vector nature of the wave packet even in the presence of unitary temporal distortions, without the need for complex time-resolved measurements.

\section{Numerical methods}
\subsection{Propagation of STVBs free space}
\begin{figure}[hpt!]
	\includegraphics[width=\linewidth]{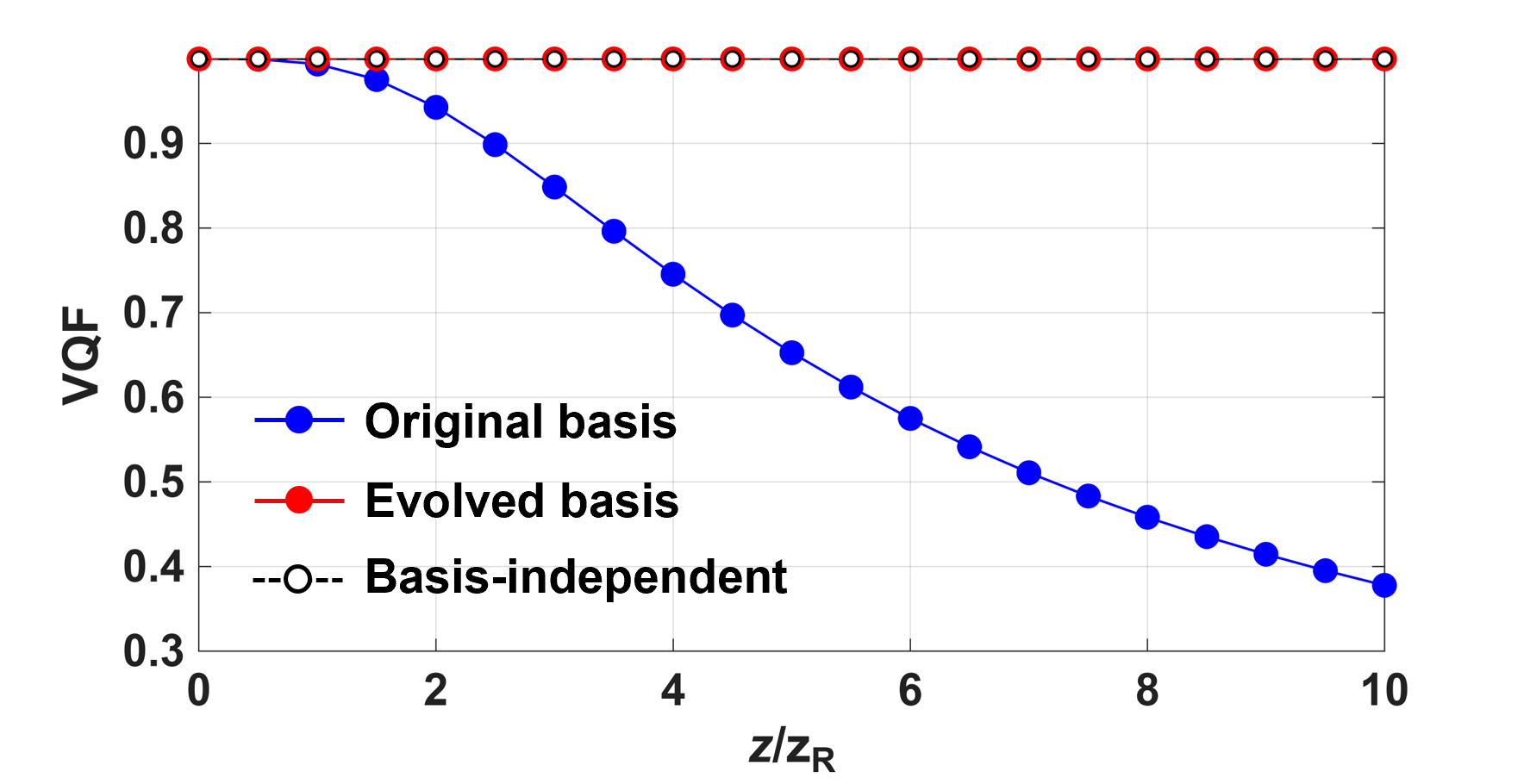}
	\caption{\textbf{VQF varying with propagation distance.}  The blue, red, and black dashed curve curves represent the $V$ values obtained without changing the spatiotemporal basis, the VQF values obtained with the spatiotemporal basis varying with propagation distance $z$, and the VQF values obtained via the
basis-independent method, respectively.}
	\label{fig:S6}
\end{figure}

\begin{figure*}[hpt!]
	\includegraphics[width=\linewidth]{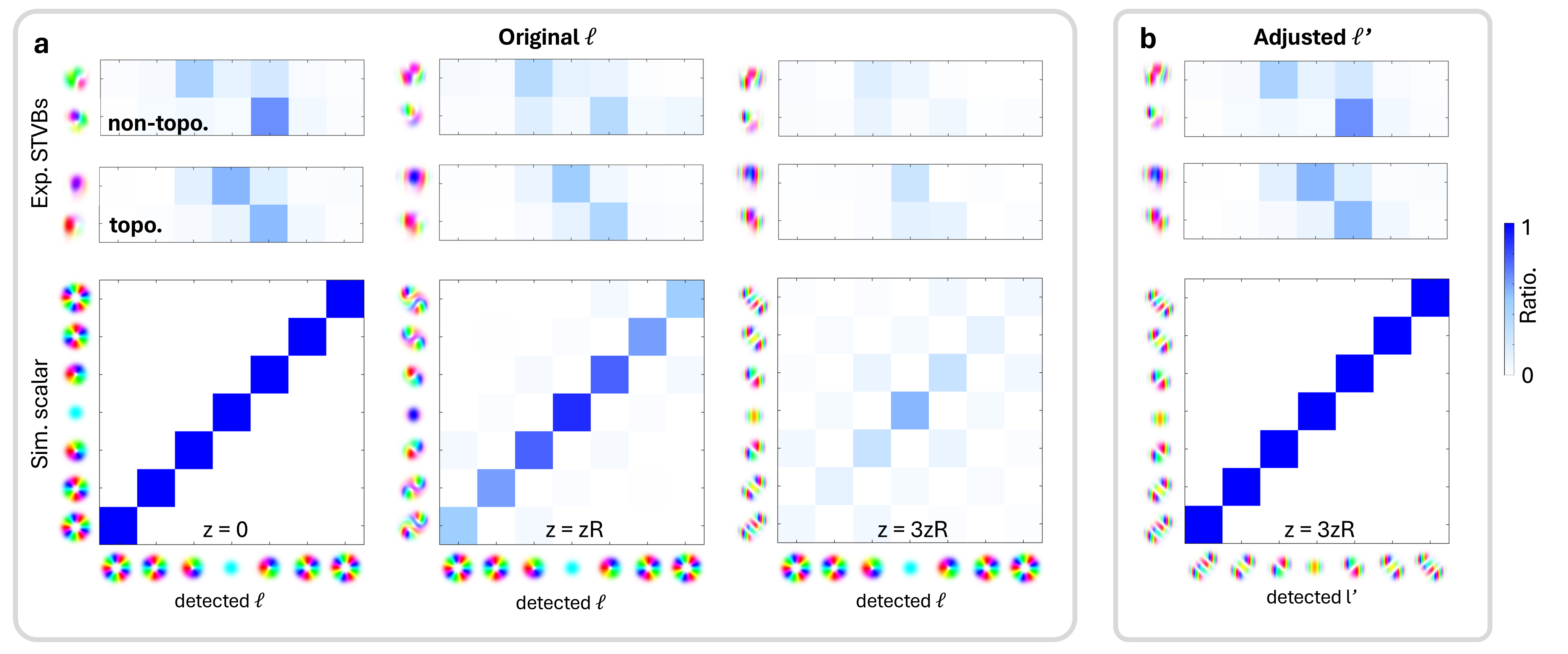}
	\caption{\textbf{Basis dependence.} \textbf{a} Crosstalk matrices at propagation distances $z = 0$, $z = z_R$, and $z = 3z_R$. The upper four rows show the projections of the measured STVB components, while the lower matrices show the projections of the corresponding ideal scalar spatiotemporal LG basis-state components. All fields are projected onto the initial spatiotemporal Laguerre-Gaussian basis functions at $z = 0$, and the color represents the normalized projection ratio. \textbf{b} Crosstalk matrix at $z = 3z_R$ after adjusting the detection basis.}
	\label{fig:VQF_scalar}
\end{figure*}

We numerically investigated the free-space propagation of a representative non-topological STVB to clarify how the VQF depends on the choice of the spatiotemporal basis. As an example, we considered an STVB in which the right- and left-handed circular polarization components are associated with two first-order ST-LG basis states carrying opposite transverse orbital angular momenta $\ell=\pm1$. The field was propagated in free space, and the VQF was evaluated using three different methods, as shown in Fig. S\ref{fig:S6}. The VQF was evaluated in three ways. First, when the initial ST-LG basis at ($z=0$) was kept fixed, the basis-dependent VQF decreased with propagation distance as shown by the blue line in Figure \ref{fig:S6}. This decay is not a loss of intrinsic polarization-spatiotemporal non-separability, but results from projection mismatch: during free-space propagation, the ST-LG basis states deform and their energy is redistributed into other basis states when projected onto the fixed initial basis. Second, when the detection basis was updated to match the propagated ST field at each $z$, the basis-dependent VQF was recovered. In the ideal case, this propagation-adapted basis completely removes the projection mismatch induced by free-space propagation, because the field is always projected onto the corresponding evolved basis states. Third, the basis-independent VQF was calculated from the global Stokes parameters. This value remains unchanged during propagation, confirming that the intrinsic non-separability of the STVB is preserved under the unitary free-space propagation channel. The different behavior of these three curves originates from the fact that ST-LG spatiotemporal vortices are not eigenfunctions of the free-space paraxial wave equation. They therefore do not propagate self-similarly in free space. The imbalance between spatial diffraction and temporal dispersion gradually deforms the initial doughnut-shaped structure in the $x-t$ plane, leading to basis-state redistribution in a fixed ST-LG projection basis. Thus, the decrease of the fixed-basis VQF should be understood as a basis-dependent projection effect, rather than degradation of the intrinsic vectorial non-separability. This is shown by the crosstalk matrices in Figure \ref{fig:VQF_scalar}\textbf{a}. The simulated initial ($z = 0$) and final ($z = z_R$) crosstalk matrices in the t-OAM basis, together with the measured polarization-resolved field components are shown. The crosstalk can be mitigated by sending in the original t-OAM basis but measuring in a new basis of distortion-adapted basis states, which is shown more explicitly for scalar spatiotemporal Laguerre-Gaussian beams at $y=0$ in Figure \ref{fig:VQF_scalar}\textbf{b} for $z = 0 $, $z_R$ and $3z_R$ for both the initial and adjusted basis. The inset beam images show the measured and detected bases in each case.

\subsection{Free space propagation parity}
The propagation dynamics and basis crosstalk behavior of spatiotemporal (ST) waves in free space are fundamentally governed by the symmetry properties of the wave equation. We define the spatiotemporal parity operator $\widehat{\mathcal{P}}$ acting on the coordinates $(x,t)$ as the simultaneous inversion of space and time, such that $\widehat{\mathcal{P}}\psi(x,t) = \psi( - x, - t)$. The spatiotemporal Laguerre-Gaussian (ST-LG) modes employed in this work are eigenstates of this parity operator, with eigenvalues determined by their topological charge $l$: $\widehat{\mathcal{P}}{LG}_{p,l} = ( - 1)^{l}{LG}_{p,l}$. Consequently, basis states with odd $l$ are odd-symmetric, while those with even $l$ are even-symmetric. Under the paraxial and narrowband approximation, the free-space transfer function
 \begin{equation}\label{prop_TF}
 H\left( k_{x},\Omega \right) = \exp\left( - i\left( \frac{k_{x}^{2}z}{2k_{0}} - \frac{\beta_{2}\Omega^{2}z}{2} \right) \right) \tag{S5}
 \end{equation}
 is an even function with respect to both transverse wavevector and frequency where the group velocity dispersion (GVD) parameter is given by $\beta_2$. Since the convolution of functions with opposite parity vanishes over a symmetric domain, the overlap integral describing the coupling between an initial basis state and a target basis state is zero if they possess opposite parity. This theoretical derivation establishes a parity selection rule: there is nearly no energy exchange (crosstalk) between ST basis subspaces of opposite parity during free-space propagation, as shown in the main text Figure 3.

 \begin{figure}[hpt!]
	\includegraphics[width=\linewidth]{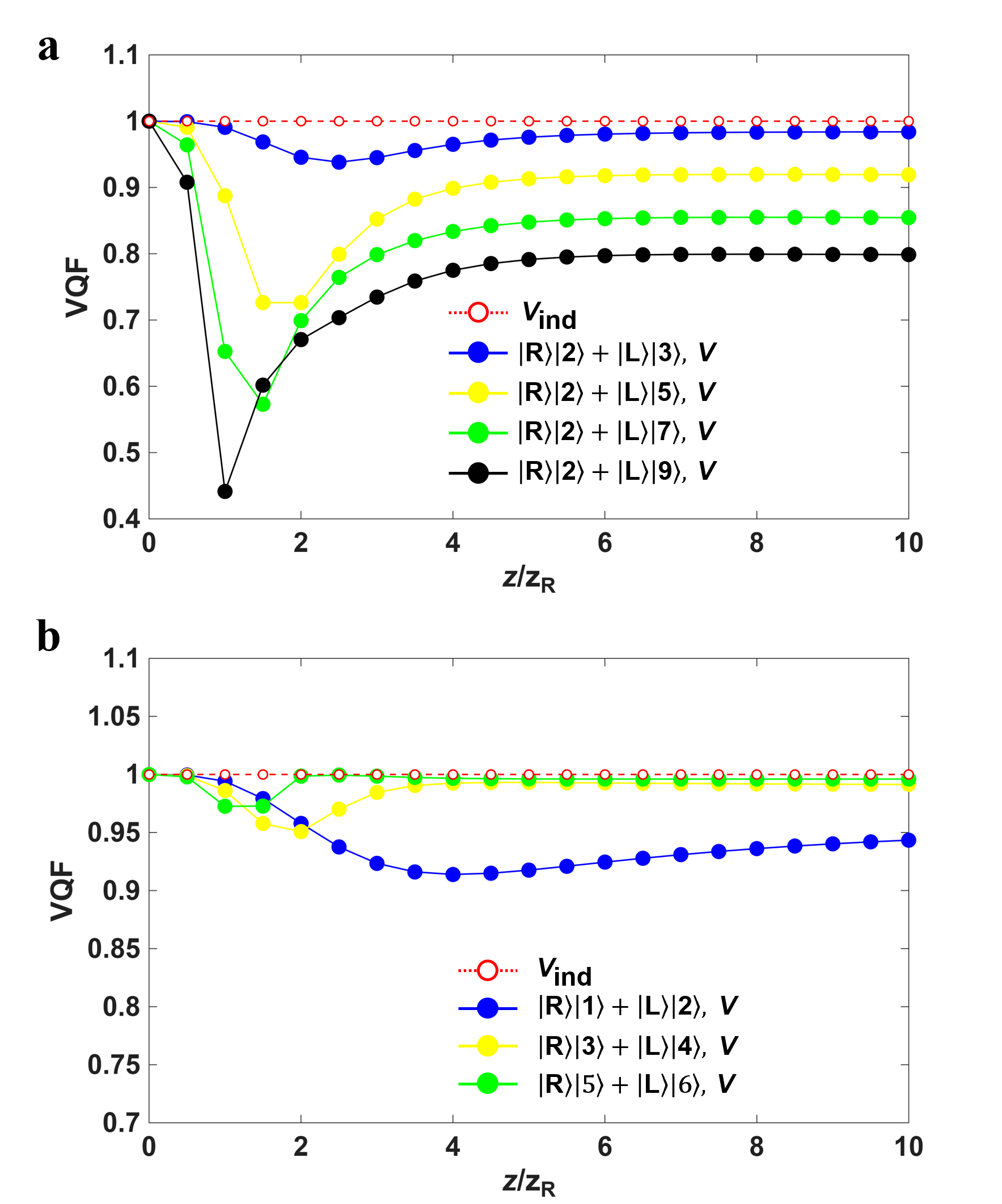}
	\caption{\textbf{Variation of the VQF with free-space propagation distance for spatiotemporal vector pulses composed of ST LG basis states with different parities.} \textbf{a} The red dashed curve, $V_{ind}$, represents the VQF obtained via the basis-independent method. The remaining curves show the basis-dependent VQFs for a configuration where the right-handed circular polarization (RCP) state corresponds to the ST ${LG}_{0,2}$ basis state, and the left-handed circular polarization (LCP) state corresponds to basis states with $l = 3,5,7,9$. \textbf{b} The red dashed curve, $V_{ind}$, again denotes the basis-independent VQF. The other curves display the basis-dependent VQFs for cases with $\delta l = 1$. Specifically, these correspond to configurations where the RCP state is associated with basis states $l = 1,3,5$, while the LCP state is associated with basis states $l = 2,4,6$, respectively.}
	\label{fig:S7}
\end{figure}
 
To investigate the implications of this rule on vector quality measurements, we performed numerical simulations on vector pulses constructed from superpositions of ST basis states with opposite parity (assigning $|R\rangle$ and $|L\rangle$ polarizations to basis states with different topological charges $l$). The results are presented in Figure \ref{fig:S7}. As a fundamental baseline, the basis-independent VQF (red dashed curves, $\text{v}_{\text{ind}}$) remains constant at unity throughout the propagation. This invariance is a direct consequence of the unitary nature of the free-space propagation channel, which preserves the global purity of the polarization subsystem regardless of the specific modal evolution or mixing details. In contrast, the basis-dependent VQF measurements (solid curves) exhibit distinct fluctuations, despite the absence of polarization crosstalk. Due to the parity selection rule derived above, the cross-coupling coefficients in the VQF calculation remain zero ($b = c = 0$). Therefore, the observed dip in VQF arises solely from the differential diffraction of the constituent modes. Although energy does not leak between the orthogonal polarization channels, the projection coefficients of the evolving fields onto the fixed measurement basis, denoted as $a(z)$ and $d(z)$, decay at unequal rates because ST-LG basis components with different $|l|$ possess distinct Gouy phase shifts and diffraction scaling laws. This creates an amplitude imbalance ($|a| \neq |d|$), causing the VQF to deviate from unity. The simulation results in Figure \ref{fig:S7}\textbf{a} illustrate that this imbalance is intensified by the disparity in mode orders: vector pulses composed of basis states with a large difference in topological charge (e.g., $l = 2$ and $l = 9$) exhibit a more pronounced difference in diffraction rates, leading to an earlier and deeper dip in the VQF curve. Furthermore, Figure \ref{fig:S7}\textbf{b} examines the case where the $l$ difference is fixed at $\Delta l = 1$. We observe that as $|l|$ increases, the VQF fluctuation diminishes. This indicates that while higher $l$ basis states diffract faster overall, the relative difference in diffraction behavior between adjacent high-order basis states (e.g., $l = 3$ and $4$) is smaller than that between low-order basis states (e.g., $l = 1$ and $2$). These findings demonstrate that even in the ideal scenario where parity prohibits crosstalk between ST basis, basis-dependent measurements remain sensitive to the differential spatiotemporal evolution of the components, reinforcing the necessity of the basis-independent toolkit for robust characterization in complex channels.

\subsection{Propagation in a diffraction-dispersion matched channel}
\begin{figure}[hpt!]
	\includegraphics[width=\linewidth]{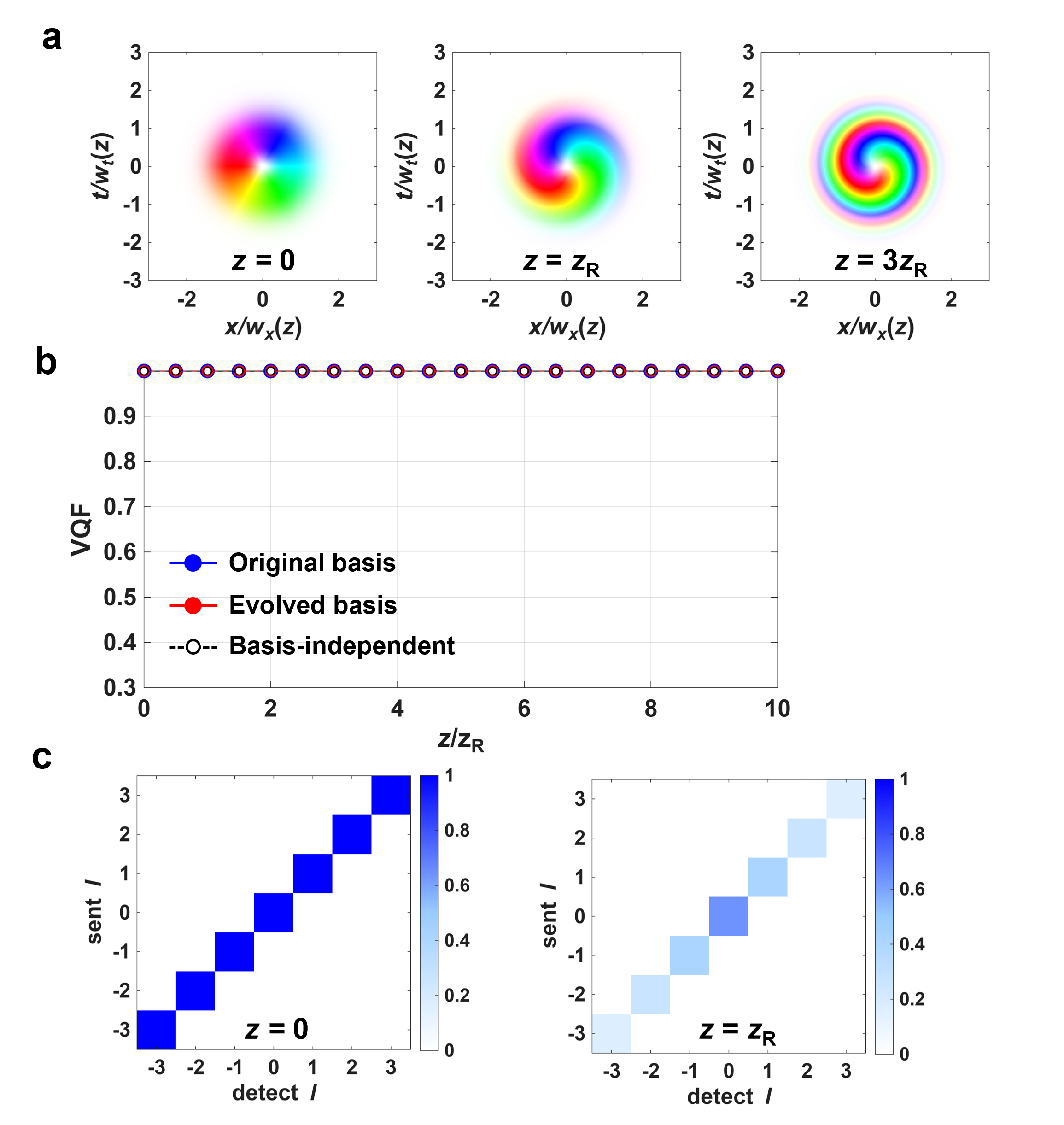}
	\caption{\textbf{Propagation in a diffraction-dispersion-balanced channel}  \textbf{a} Propagation of an ST-LG wave packet with $\ell = 1$ in an ideal diffraction-dispersion-balanced medium. \textbf{b} VQF curves of the ideal spatiotemporal vector wave packet. The blue, red, and black curves represent the VQF obtained without changing the spatiotemporal basis (original), the VQF obtained using a $z$-varying spatiotemporal basis (evolved), and the VQF obtained via the basis-independent method, respectively. \textbf{c} Crosstalk matrices of the ST-LG basis states at $z = 0$ and $z = z_R$. The rows correspond to the launched ST-LG basis states after propagation, and the columns correspond to the fixed detection basis defined at $z=0$. The color scale indicates the magnitude of the crosstalk coefficients between different basis states.
 }
	\label{fig:S8}
\end{figure}

To conclusively demonstrate that the degradation of basis-dependent VQF in free space stems from the spatiotemporal asymmetry (i.e., diffraction without dispersion), we investigated the pulse propagation in a theoretical "diffraction-dispersion balanced" medium. In this channel, the group velocity dispersion (GVD) parameter $\beta_{2}$ is tuned to satisfy the condition $\beta_{2} = -{w_{t}^{2}}/({n(\omega_{0})k_{0}w_{x}^{2}})$, where $w_{t}$ and $w_{x}$ denote the temporal and spatial widths of the pulse respectively. This condition renders the temporal evolution operator mathematically isomorphic to the spatial diffraction operator, thereby restoring spatiotemporal symmetry and allowing the ST-LG wave packets to propagate as a shape-invariant eigenmode. Figure \ref{fig:S8}\textbf{a} visualizes the propagation of an ideal $l = 1$ ST-LG wave packet in this balanced medium. In stark contrast to the free-space case, the intensity profile maintains its characteristic doughnut topology throughout the propagation distance from $z = 0$ to $3z_{R}$. The pulse undergoes uniform scaling due to simultaneous diffraction and dispersion, but its structural integrity remains intact without astigmatic deformation. The impact of this symmetry restoration on the vector quality is quantified in Figure \ref{fig:S8}\textbf{b}. The VQF curves obtained via three different methods: basis-independent (black dashed), fixed-basis (red), and adaptive-basis (blue). All of them remain nearly flat and close to unity (VQF $\approx 1$) over a long propagation distance of $10z_{R}$, which confirms that when the modal shape is preserved, the amplitude imbalance responsible for the VQF dip in free space is effectively suppressed. Furthermore, we analyzed the modal purity during propagation by calculating the crosstalk matrix between the ideal propagated ST-LG wave packets at $z = z_{R}$ and the original basis modes at $z = 0$, as shown in Figure \ref{fig:S8}\textbf{c}. As expected, the matrix is strictly diagonal, indicating that there is no crosstalk.

\section{Extracting the Skyrmion wrapping number}
\begin{figure}[h!]
	\includegraphics[width=\linewidth]{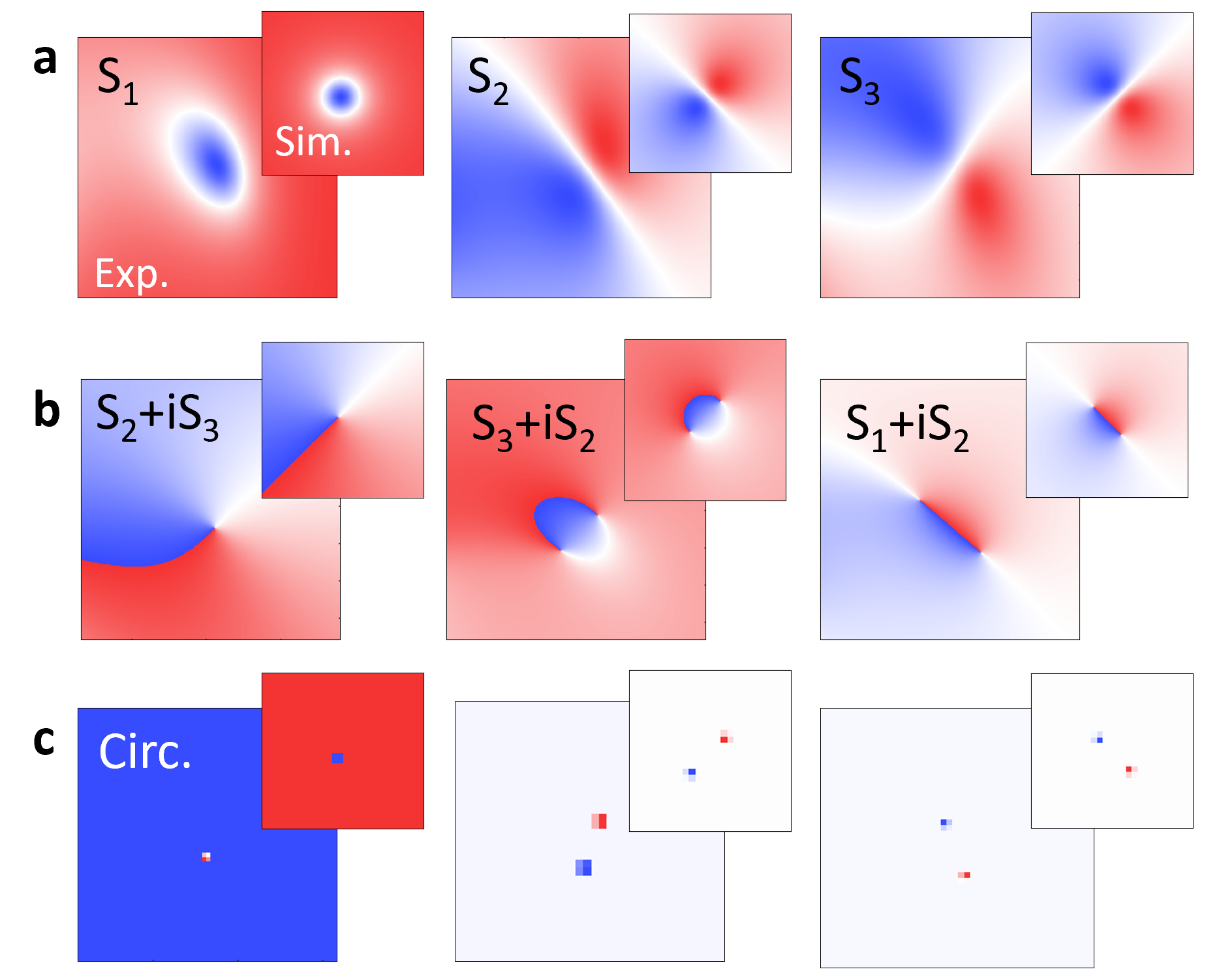}
	\caption{\textbf{Measuring topology with line integral for ideal cases.} \textbf{a}-\textbf{c} $N=1$ case with simulation and experiment. \textbf{a} Locally renormalised Stokes parameters. \textbf{b} Polarisation phases. \textbf{c} Circulation computed after an applied Gaussian filter and thresholding of $2\%$. Relative phases which do not change $N$ have been added to the simulation case to match the experiment.} 
	\label{fig:S5}
\end{figure}

From the full field information we can compute the the Stokes parameters as
\begin{equation}
    \begin{split}
        s_0 &= I_H(x,\tau) + I_V(x,\tau),\\
        s_1 &= I_H(x,\tau) - I_V(x,\tau),\\
        s_2 &= I_D(x,\tau) - I_A(x,\tau) \\
        s_3 &=  I_R(x,\tau) - I_L(x,\tau),\\
    \end{split}
\end{equation}
using the intensities for horizontal, vertical, diagonal, anti-diagonal, right and left circularly polarised light in the $y=0$ plane. The locally renormalised stokes vector is then given by 
$\mathbf{S}(x,\tau)=[s_1(x,\tau)\ s_2(x,\tau)\ s_3(x,\tau)]/s_0(x,\tau).$
where we explicitly write the the time degree of freedom, $\tau$ to emphasize the difference to traditional spatial Stokes skyrmions in the $x-y$ plane. These are depicted in Figure \ref{fig:S5}\textbf{a} for the $N=1$ case. Next we define the mapping to the spatio-temporal Poincare sphere as in \cite{pires2025structuring} where the spatiotemporal field wraps the Poincare sphere $N$ times where the mapping is given by
$(x,\tau)\in \mathcal{R}^2 \rightarrow \mathcal{S}^2,$
where $\mathcal{R}^2$ is the region of the reconstructed field over which the topology is evaluated and $\mathcal{S}^2 $ is the spatio-temporal  Poincare sphere with unit radius such that every point in  $\mathcal{R}^2$ maps to $\mathcal{S}^2 $. The skyrmion number $N$ is calculated using a  line integral approach where the skyrmion number is given by 
\begin{equation}
    N = \frac{1}{2}\left ( \sum_j S_z^{(j)}N_j + S_z^{(\infty)} N_\infty \right )
    \label{eq:skyrme_calc}
\end{equation}
where $S_z^{(\infty)} = S_z^{(\infty)}(x,\tau)$ is the locally renormalised Stokes vector at infinity and  $S_z^{(j)} = S_z^{(j)}(x,\tau)$ is the value of the Stokes parameter at position $j$ in the $x-t$ plane of the $y=0$ slice of the pulse. Here $N_j$ represents the charge of the phase singularity $j$ which is computed as in \cite{Peters2026extracting} but now in the $x-t$ plane instead of $x-y$. The polarisation singularities used in Equation \ref{eq:skyrme_calc} are computed from the polarisation phases shown in Figure \ref{fig:S5}\textbf{b} and \textbf{c}.

\end{document}